\documentclass[preprint,aps,nofootinbib]{revtex4-1}
\pdfoutput=1

\usepackage{dcolumn}
\usepackage{bm}
\usepackage{epsfig}
\usepackage{graphicx}
\usepackage{float}
\usepackage{paralist}

\usepackage{amsmath}
\usepackage{amsfonts}
\usepackage{amssymb}
\usepackage{adjustbox}
\usepackage{color}
\usepackage{xcolor}
\usepackage{adjustbox}
\usepackage{subcaption}
\usepackage[toc,page]{appendix}

\usepackage[utf8]{inputenc}
\usepackage{slashed}
\usepackage{indentfirst}
\usepackage{caption}
\usepackage{subcaption}
\usepackage{hyperref} 
\usepackage{natbib} 
\usepackage{ulem}
\usepackage{braket}
\usepackage{ulem}

\def\tt{\hat{\tau}}
\def\mtt{m_{\hat{\tau}}}
\def\tn{\hat{\nu}}
\def\mtn{m_{\hat{\nu}}}
\def\gd{\hat{\gamma}}
\def\mgd{m_{\hat{\gamma}}}
\def\ttB{{\tau}_B}
\def\tnB{{\nu}_B}
\def\mttB{m_{{\tau}_B}}
\def\mtnB{m_{{\nu}_B}}
\def\ttBR{{\tau}_{B,R}}
\def\tnBR{{\nu}_{B,R}}
\def\ttBL{{\tau}_{B,L}}
\def\tnBL{{\nu}_{B,L}}
\def\y{\hat{e}}

\begin{document}

\title{Coscattering/Coannihilation Dark Matter in a Fraternal Twin Higgs Model
}
\author{Hsin-Chia Cheng$^{a,b}$, Lingfeng Li$^{a}$, Rui Zheng$^{a}$}
\affiliation{$^a$Department of Physics, University of California Davis, Davis, California 95616, USA\\
$^b$School of Natural Sciences, Institute for Advanced Study, Princeton, New Jersey 08540, USA}

\begin{abstract}
Dark matter candidates arise naturally in many models that address the hierarchy problem. In the fraternal twin Higgs model which could explain the absence of the new physics signals at the Large Hadron Collider (LHC), there are several viable dark matter candidates. In this paper we study the twin neutrino in the mass range $\sim$ 0.1--10 GeV as the dark matter. The thermal relic density is determined by the interplay of several annihilation and scattering processes between the twin neutrino, twin tau, and twin photon, depending on the order of the freeze-out temperatures of these processes. Besides the common coannihilation scenario where the relic density is controlled by the twin tau annihilation, it can realize the recently discovered coscattering phase if the scattering of the twin neutrino into the twin tau freezes out earlier than the twin tau annihilation. We also provide a method to calculate the thermal relic density in the intermediate regime where both coannihilation and coscattering processes contribute to the determination of the dark matter density.  We show that the right amount of dark matter can be obtained in various scenarios in different regions of the parameter space. The current experimental constraints and future probes into the parameter space from direct detections, cosmological and astrophysical bounds, dark photon searches, and displaced decays at colliders, are discussed.
\end{abstract}
\maketitle

\section{Introduction}
\label{sec:introduction}

The hierarchy problem and the dark matter are two main motivations for new physics near the electroweak (EW) scale. In the standard model (SM), the Higgs field receives large quadratically divergent contributions to its potential from the interactions with SM particles, in particular, the top quark and weak gauge bosons. For a natural EW symmetry breaking scale, new particles are expected to be close to the EW scale to cut off these quadratic contributions. On the other hand, a stable weakly interacting massive particle (WIMP) with a mass around the EW scale gives a right amount of thermal relic from the Hot Big Bang to account for the dark matter in the universe. It is called ``WIMP miracle.'' Such a dark matter particle candidate also often appears naturally in models which address the hierarchy problem. The most popular and most studied examples are the supersymmetric (SUSY) extensions of SM. With a conserved $R$-parity, the lightest neutralino is stable and represents a good dark matter candidate. The models that can explain both the hierarchy problem and dark matter are particularly attractive because they provide a link between the two mysterious problems.

The new particles related to the hierarchy problem and the WIMPs have been extensively searched at various experiments. So far none of them has been discovered. The LHC has put very strong bounds on new colored particles that can cancel the SM top loop contribution to the Higgs mass. Except for some special cases, the bounds on the masses of new colored particles generically exceed 1 TeV. This would imply a quite severe tuning of the Higgs mass if the top loop is not canceled below 1 TeV. Direct searches of DM also put strong bounds on the scattering cross sections of the DM particle with nucleons. A big fraction of the expected region of the WIMP parameter space from typical SUSY models is excluded, though there are still surviving scenarios. These null experimental results have prompted people to wonder that the standard pictures such as SUSY might not be realized at the electroweak scale in nature. Alternative solutions to the hierarchy problem and DM where the interactions between new particles and SM particles are stealthier should be taken more seriously.

For the hierarchy problem, the ``neutral naturalness'' models gained increasing attentions in recent years. In these models, the top quark partners which regularize the top loop contribution to the Higgs mass do not carry SM color quantum numbers, and hence are not subject to the strong bounds from the LHC. The mirror twin Higgs model~\cite{Chacko:2005pe} is the first example and is probably the stealthiest one. The twin sector particles are SM singlets but charged under their own $SU(3) \times SU(2) [\times U(1)]$ gauge group. They are related to the SM sector by a $Z_2$ symmetry. As a result, the mass terms of the Higgs fields of the SM and twin sectors exhibit an enhanced $SU(4)$ symmetry. The 125 GeV Higgs boson arises as a pseudo-Nambu-Goldstone boson (PNGB) of the spontaneously broken $SU(4)$ symmetry. The twin sector particles are difficult to produce at colliders because they do not couple to SM gauge fields. The main experimental constraints come from the mixing between the SM Higgs and the twin Higgs, which are rather weak. The model can still be natural without violating current experimental bounds.

The next question is whether the neutral naturalness models like the twin Higgs possess good dark matter candidates. In the fraternal version of the twin Higgs model~\cite{Craig:2015pha}, people have shown that there are several possible dark matter candidates~\cite{Garcia:2015loa,Craig:2015xla,Garcia:2015toa,Farina:2015uea}. The fraternal twin Higgs takes a minimal approach in addressing the hierarchy problem using the twin Higgs mechanism. In this model, the twin fermion sector only contains the twin partners of the third generation SM fermions, since only the top Yukawa coupling gives a large contribution to the Higgs mass that needs to be regularized below the TeV scale. The twin $U(1)$ gauge boson can be absent or can have a mass without affecting the naturalness. The fraternal twin Higgs model can avoid potential cosmological problems of an exact mirror twin Higgs model which contains many light or massless particles in the twin sector. Refs.~\cite{Garcia:2015loa,Craig:2015xla} showed that the twin tau can be a viable dark matter candidate. The correct relic density is obtained for a twin tau mass in the range of 50--150 GeV, depending on other model parameters. If twin hypercharge is gauged, then the preferred mass is lighter, in the 1--20 GeV range~\cite{Craig:2015xla}. Another possibility is asymmetric dark matter from the twin baryon made of twin $b$-quarks, where the relic density is set by the baryon asymmetry in the twin sector~\cite{Garcia:2015toa,Farina:2015uea}. 

In this paper, we explore a new scenario where the dark matter is the twin neutrino, in the mass range $\sim$ 0.1--10~GeV. In previous studies of twin tau dark matter, its stability is protected by the twin $U(1)_\text{EM}$ symmetry, which is assumed to be a good symmetry, either gauged or global. Here we consider that the twin $U(1)_\text{EM}$ is broken so that the twin photon acquires a mass to avoid potential cosmological problems. In this case, the twin tau and the twin neutrino can mix so the twin tau can decay to the twin neutrino if the twin tau is heavier. The twin neutrino, on the other hand, being the lightest twin fermion, can be stable due to the conservation of the  twin lepton number or twin lepton parity.  An interesting scenario is that if the twin photon, twin neutrino, and twin tau all have masses of the same order around a few GeV or below, the right amount of dark matter relic density from twin neutrinos can be obtained. The relic density is controlled by the coannihilation and recently discovered coscattering processes~\cite{DAgnolo:2017dbv,Garny:2017rxs}.

The coscattering phase is considered as the fourth exception in the calculation of thermal relic abundances in addition to the three classical cases enumerated in Ref.~\cite{Griest:1990kh}. It is closely related to the coannihilation case as both require another state with mass not far from the dark matter mass so that the partner state can play an important role during decoupling. The difference is that in the coannihilation phase the relic density is controlled by the freeze-out of the annihilation processes of these particles, while in coscattering phase the relic density is controlled by the freeze-out of the inelastic scattering of a dark matter particle into the partner state. Because of the energy threshold of the upward scattering, the coscattering process has a strong momentum dependence. This makes the relic density calculation quite complicated. The standard DM calculation tools such as micrOMEGAs~\cite{Belanger:2018ccd}, DarkSUSY~\cite{Bringmann:2018lay}, and MadDM~\cite{Ambrogi:2018jqj} do not apply and one needs to solve the momentum-dependent Boltzmann equations. Also, because of the momentum dependence of the coscattering process, we find that there are parameter regions of mixed phase, i.e., the relic abundance is controlled partially by coannihilation and partially by coscattering. We investigate in detail the relevant parameter space and perform calculations of the relic abundance in different phases, including situations where it is controlled by coannihilation, by coscattering, or by both processes. The calculation in the mixed phase is more involved and we discuss a relatively simple method to obtain the DM abundance with good accuracies.

The twin neutrino DM does not couple to SM directly.  Its couplings to matter through mixings of the photons or the Higgses between the SM sector and the twin sector are suppressed, so the direct detection experiments have limited sensitivities. Some of the main constraints come from indirect detections and searches of other associated particles. Its annihilation through twin photon is constrained by Cosmic Microwave Background (CMB), 21cm line absorption,  and Fermi-LAT data. For associated particles, the light dark photon searches provide some important constraints and future probes.

The paper is organized as follows. In Sec.~\ref{sec:FTH} we first give a brief summary of the fraternal twin Higgs model and its possible DM candidates. Then we focus the discussion on the sector of our DM scenario, i.e., the twin neutrino as the DM, and its coannihilation/coscattering partners, the twin tau and the twin photon. In Sec.~\ref{sec:Kinematics} we enumerate the relevant processes and discuss their roles in controlling the DM abundance in different scenarios.  Sec.~\ref{sec:calculation} describes how to evaluate the DM relic density in different phases, including the coannihilation phase, the coscattering phase, and the mixed phase. Some details about the calculations are collected in the Appendices. Our numerical results for some benchmark models are presented in Sec.~\ref{sec:numerical}. In Sec.~\ref{sec:constraints} we discuss various experimental constraints and future probes of this DM scenario. The conclusions are drawn in Sec.~\ref{sec:conclusions}.

\section{Fraternal Twin Higgs and Light DM}
\label{sec:FTH}

The twin Higgs model postulates a mirror (or twin) sector which is related to the SM sector by a $Z_2$ symmetry. The particles in the twin sector are completely neutral under the SM gauge group but charged under the twin $SU(3)\times SU(2) \times U(1)$ gauge group. Due to the $Z_2$ symmetry, the Higgs fields of the SM sector and the twin sector exhibit an approximate $U(4)$ (or $O(8)$) symmetry. The $U(4)$ symmetry is spontaneously broken down to $U(3)$ by the Higgs vacuum expectation values (VEVs). A phenomenological viable model requires that the twin Higgs VEV $f$ to be much larger than the SM Higgs VEV $v$, $f/v \gtrsim 3$, so that the light uneaten pseudo-Nambu-Goldstone boson (pNGB) is an SM-like Higgs boson.  (The other six Nambu-Goldstone bosons are eaten and become the longitudinal modes of the $W,\, Z$ bosons of the SM sector and the twin sector.) This requires a small breaking of the $Z_2$ symmetry. The one-loop quadratically divergent contribution from the SM particles to the Higgs potential is cancelled by the twin sector particles, which are heavier than their SM counterparts by the factor of $f/v$.  The model can be relatively natural for $f/v \sim 3-5$.

The twin sector particles are not charged under the SM gauge group so it is difficult to look for them at colliders. However, if there is an exact mirror content of the SM sector and the couplings respect the $Z_2$ symmetry, there will be light particles (photon, electron, neutrinos) in the twin sector. They can cause cosmological problems by giving a too big contribution to $N_\text{eff}$.\footnote{Some solutions within the mirror twin Higgs framework can be found in Refs.~\cite{Farina:2015uea,Chacko:2016hvu,Craig:2016lyx,Barbieri:2016zxn,Csaki:2017spo,Barbieri:2017opf}.} In addition, in general one expects a kinetic mixing between two $U(1)$ gauge fields. If the twin photon is massless, its kinetic mixing with the SM photon is strongly constrained. On the other hand, these light particles have small couplings to the Higgs and hence play no important roles in the hierarchy problem. One can take a minimal approach to avoid these light particles by only requiring the $Z_2$ symmetry on the parts which are most relevant for the hierarchy problem. This is the fraternal twin Higgs (FTH) model proposed in Ref.~\cite{Craig:2015pha}. The twin sector of the FTH model can be summarized below.

\begin{itemize}
\item The twin $SU(2)$ and $SU(3)$ gauge couplings should be approximately equal to the SM $SU(2)$ and $SU(3)$ gauge couplings. The twin hypercharge does not need to be gauged. If it is gauged, its coupling can be different from the SM hypercharge coupling, as long as it is not too big to significantly affect the Higgs potential. Also, the twin photon can be massive by spontaneously breaking the $U(1)$ gauge symmetry or simply writing down a Stueckelberg mass term.

\item There is a twin Higgs doublet. Together with the SM Higgs doublet there is an approximate $U(4)$-invariant potential. The twin Higgs doublet acquires a VEV $f\gg v$, giving masses to the twin weak gauge bosons and twin fermions.

\item The twin fermion sector contains the third generation fermions only. The twin top Yukawa coupling needs to be equal to the SM top Yukawa coupling to a very good approximation so that their contributions to the Higgs potential can cancel. The twin bottom and twin leptons are required for anomaly cancellation, but their Yukawa couplings do not need to be equal to the corresponding ones in the SM, as long as they are small enough to not generate a big contribution to the Higgs potential.

\end{itemize}

The collider phenomenology of the FTH model mainly relies on the mixing of the SM and twin Higgs fields. In typical range of the parameter space, one often expects displaced decays that constitute an interesting experimental signature. Here we focus on the DM. A natural candidate is the twin tau. Since its Yukawa coupling needs not to be related to the SM tau Yukawa coupling, its mass can be treated as a free parameter. It is found that a right amount of thermal relic abundance can be obtained for a twin tau mass in the range of 50--150 GeV if the twin hypercharge is not gauged~\cite{Garcia:2015loa,Craig:2015xla}. It corresponds to a twin tau Yukawa coupling much larger than the SM tau Yukawa coupling. The requirement that the twin tau Yukawa coupling does not reintroduce the hierarchy problem puts a upper limit $\sim 200$ GeV on the twin tau mass. A twin tau lighter than $\sim 50$ GeV would generate an overabundance which overcloses the universe. The relic density can be greatly reduced if a light twin photon also exists, because it provides additional annihilation channels for the twin tau. If the twin photon coupling strength is similar to the SM photon coupling, the annihilation will be too efficient and it will be difficult to obtain enough DM. For a twin photon coupling $\hat{\alpha} \sim 0.03\, \alpha_\text{EM}$, a right amount of relic density can be obtained for a twin tau mass in the range of 1--20 GeV~\cite{Craig:2015xla}.

In the twin tau DM discussion, its stability is assumed to be protected by the twin $U(1)_\text{EM}$ symmetry. However, if twin $U(1)_\text{EM}$ is broken and the twin photon has a mass, the twin tau may be unstable and could decay to the twin neutrino if the twin neutrino is lighter. This is because that the twin tau and the twin neutrino can mix due to the twin $U(1)_\text{EM}$ breaking effect. On the other hand, if the twin lepton number (or parity) is still a good symmetry, the lightest state that carries the twin lepton number (parity) will be stable. In this paper we will assume that it is the twin neutrino and consider its possibility of being the DM.

\subsection{Twin lepton mixings}

If the twin $U(1)_\text{EM}$ (or equivalently twin hypercharge) is broken, it is possible to write down various Dirac and Majorana masses between the left-handed and right-handed twin tau and twin neutrino fields. For simplicity, we consider the case where the twin lepton number remains a good symmetry, which can be responsible for the stability of DM. This forbids the Majorana mass terms. 

The twin tau and twin neutrino receive the usual Dirac masses from the twin Higgs VEV,
\begin{eqnarray}
-\mathcal{L}& \supset & y_{\ttB} \, {L}_B \, \widetilde{H}_B \,\ttBR^c + y_{\tnB}\, {L}_B \, {H}_B\, \tnBR^c + \text{h.c.} \nonumber \\
&\supset & \frac{y_{\ttB} f}{\sqrt{2}}\, \ttBL \ttBR^c + \frac{y_{\tnB} f}{\sqrt{2}}\, \tnBL \tnBR^c + \text{h.c.},
\label{eq:diracmass}
\end{eqnarray}
where the subscript $B$ represents the twin sector fields, and $\widetilde{H}_B = i\sigma_2 H^\ast_B$ transforms as $(\bf{3}, \bf{2})_{-1/2}$ under the twin gauge group. The twin hypercharge breaking can be parameterized by a spurion field $S$ which is a singlet under $SU(3)_B \times SU(2)_B$ but carries $+1$ twin hypercharge (also $+1$ twin electric charge).  It can come from a VEV of a scalar field which breaks $U(1)_B$ spontaneously. The radial component is assumed to be heavier than the relevant particles (twin photon, tau, and neutrino) here and plays no role in the following discussion. Using the spurion we can write down the following additional lepton-number conserving mass terms,
\begin{eqnarray}
-\mathcal{L}\supset\frac{ d_1}{\Lambda} S {L}_B \widetilde{H}_B \tnBR^c + \frac{d_2}{\Lambda}  S^\dagger {L}_B {H}_B \ttBR^c \, .
\label{eq:mixingmass}
\end{eqnarray}
The mass matrix of the twin tau and twin neutrino is then given by
\begin{eqnarray}
(\begin{array}{cc}
\ttBR^c & \tnBR^c\\ 
\end{array})
\left(\begin{array}{cc}
\mttB & \mu_{2}\\ 
\mu_{1} & \mtnB
\end{array}\right)
\left(\begin{array}{c}
\ttBL \\
\tnBL
\end{array}\right),
\label{eqn:mixing 1}
\end{eqnarray}
where 
\begin{eqnarray}
\mttB = \frac{y_{\hat{\tau}} f}{\sqrt{2}}, \quad \mtnB = \frac{y_{\hat{\nu}} f}{\sqrt{2}}, \quad \mu_1 = \frac{d_1 f S}{\sqrt{2}\Lambda}, \quad  \mu_2 = \frac{d_2 f S }{\sqrt{2}\Lambda}.
\end{eqnarray}

The mass matrix can be diagonalized by the rotations
\begin{eqnarray}
\left(\begin{array}{c}
\hat{\tau}_R^{c}\\
\hat{\nu}_R^{c}
\end{array}\right)=\left(\begin{array}{cc}
\cos\theta_1 &\sin\theta_1\\ 
-\sin\theta_1 & \cos\theta_1
\end{array}\right)
\left(\begin{array}{c}
\ttBR^{c}\\
\tnBR^{c}
\end{array}\right),\quad
\left(\begin{array}{c}
\hat{\tau}_L^{} \\
\hat{\nu}_L^{}
\end{array}\right)=\left(\begin{array}{cc}
\cos\theta_2 &\sin\theta_2\\ 
-\sin\theta_2 & \cos\theta_2
\end{array}\right)
\left(\begin{array}{c}
\ttBL^{} \\
\tnBL^{}
\end{array}\right) \, ,\label{eqn:thetamatrix}
\end{eqnarray}
where the mass eigenstates in the twin sector are labelled with a hat (\,$\hat{}\,$).
We assume that the off-diagonal masses $|\mu_1|, |\mu_2| \ll  \mttB - \mtnB$ so that the mixing angles $\theta_1, \theta_2$ are small. This is reasonable given that $\mu_1, \mu_2$ arise from higher dimensional operators and require an insertion of the twin hypercharge breaking VEV, which is assumed to be small for a light twin photon. For our analysis, to reduce the number of independent parameters, we further assume that one of the off-diagonal masses dominates, i.e., $\mu_1 \gg \mu_2$, so that we can ignore $\mu_2$. In this case, we obtain two Dirac mass eigenstates $\tt$ and $\tn$ which are labeled by their dominant components. The two mass eigenvalues are
\begin{equation}
m^2_{\tt,\tn}=\frac{1}{2} \left(\mu_{1}^2+\mttB^2+\mtnB^2\pm\sqrt{\left(\mu_{1}^2+\mttB^2+\mtnB^2\right)^2-4 \mttB^2 \mtnB^2}\right),
\end{equation}
and the two mixing angles are given by
\begin{equation}
\sin\theta_1 = \frac{\mu_{1} \mttB}{\mttB^2 -\mtnB^2 +\mu_{1}^2}, \quad \sin\theta_2=\frac{\mtnB}{\mttB}\sin\theta_1\, ,
\end{equation}
in the small mixing angle limit. There is no qualitative difference in our result if $\mu_1$ and $\mu_2$ are comparable except that the two mixing angles become independent.

We are interested in the region of parameter space where the twin tau $\tt$, twin neutrino $\tn$, and twin photon $\gd$ have masses of the same order in the range $\sim 0.1-10$~GeV, with $\mtt > \mtn > \mgd$. Compared with Ref.~\cite{DAgnolo:2017dbv}, the twin neutrino $\tn$ plays the role of $\chi$ which is the DM, $\tt$ corresponds to $\psi$, the coannihilation/coscattering partner of the DM particle, and $\gd$ corresponds to the mediator $\phi$. Following Ref.~\cite{DAgnolo:2017dbv}, we define two dimensionless parameters,
\begin{equation}
r \equiv \frac{\mgd}{\mtn}, \qquad \Delta \equiv \frac{\mtt-\mtn}{\mtn},
\end{equation}
which are convenient for our discussion. The region of interest has $r < 1$ and $0< \Delta \lesssim 1$.

The mass spectrum and mixing pattern would be more complicated if Majorana masses for the twin leptons are allowed. In addition to the standard Majorana mass for the right-handed twin neutrino, all other possible terms can arise from higher dimensional operators with insertions of the spurion field $S$ (and the twin Higgs field $H_B$), filling the $4\times 4$ mass matrix of $(\ttBL, \tnBL, \ttBR^c, \tnBR^c)$.  There are four mass eigenstates and many more mixing angles. The stability of the lightest eigenstate can be protected by the twin lepton parity in this case. Because the twin photon couples off-diagonally to Majorana fermions, the analysis of annihilation and scattering needs to include all four fermion eigenstates, which becomes quite complicated. Nevertheless, one can expect that there are regions of parameter space where the correct relic abundance can be obtained through coannihilation and/or coscattering processes in a similar way to the case studied in this work.

\section{Relevant Processes for the Thermal Dark Matter Abundance}
\label{sec:Kinematics}

At high temperature, the SM sector and the twin sector stay in thermal equilibrium through the interactions due to the Higgs mixing and the kinetic mixing of the $U(1)$ gauge fields. As the universe expands, the heavy species in the twin sector decouple from the thermal bath and only the light species including twin photon ($\gd$), twin tau ($\tt$), and twin neutrino ($\tn$) survive. These light species talk to SM mainly via the kinetic mixing term, $-(\epsilon/2) F_{\mu\nu}\hat{F}^{\mu\nu}$, between $\gamma$ and $\gd$. We assume that the kinetic mixing is big enough ($ \epsilon \gtrsim 10^{-9}$) to keep the twin photon in thermal equilibrium by scattering off light SM leptons~\cite{DAgnolo:2017dbv} during the DM freeze-out. The DM abundance is then controlled by several annihilation and scattering processes.

{\bf Annihilation processes}:
\begin{equation*}
\tn\tn \to \gd\gd~(\hm{A}), \qquad
\tt\tn\to \gd\gd~(\hm{C_A}), \qquad 
\tt\tt\to \gd\gd~( \hm{C_S}).
\end{equation*}
The coupling of $\tn$ to $\gd$ arises from mixing with the twin tau. As a result, for small mixings the usual annihilation process $\hm{A}$ for the $\tn$ DM is suppressed by $\theta_1^4$, while the coannihilation process $\hm{C_A}$ (where the subscript $\hm{A}$ stands for asymmetric) is suppressed by $\theta_1^2$. There is no mixing angle suppression for the coannihilation process $\hm{C_S}$ (where the subscript $\hm{S}$ represents symmetric or sterile). On the other hand, the Boltzmann suppression goes the other way, the Boltzmann factors for the three processes are $\sim e^{-2\mtn/T}$, $e^{-(\mtn+\mtt)/T}$, and $e^{-2\mtt/T}$ respectively. 

{\bf (Co)scattering process}:
\begin{equation*}
\tn\gd\to \tt\gd~(\hm{S}).
\end{equation*}
It is suppressed by $\theta_1^2$. Because $\tt$ is assumed to be heavier than $\tn$, the initial states particles $\tn$ and $\gd$ must carry enough momenta for this process to happen. The coscattering process therefore has a strong momentum dependence. Ignoring the momentum dependence for a moment, the Boltzmann factor can be estimated to be $\sim e^{-(\mtt+\mgd)/T}$.

In addition, there is also the decay process 
\begin{equation*}
\tt \to \tn\gd^{(\ast)}~(\hm{D}).
\end{equation*} 
On-shell decay only occurs if $\mtt > \mtn + \mgd\; ( \Delta > r)$. In this case the inverse decay ($\hm{ID}$) plays a similar role as the coscattering process since both convert $\tn$ to $\tt$, but the rate is much larger. It turns out that if the (inverse) decay is open, the relic abundance is simply determined by the coannihilation because the inverse decay process decouples later.
The majority of the parameter space we focus on has $\mtt< \mtn+\mgd\; (\Delta < r)$. In this case, the twin photon has to be off-shell then decays to SM fermions. It is further suppressed by $\epsilon^2$ so it can be ignored  during the freeze-out. It is however responsible for converting the remaining $\tt$ to $\tn$ eventually after the freeze-out.

If the mixing is large and/or $\Delta$ is large so that $\theta_1^2 e^{-2\mtn/T} > e^{-(\mtt+\mtn)/T}$ during freeze-out, then the annihilation process $\hm{A}$ will dominate and we will have the usual WIMP scenario. However, for such a WIMP DM lighter than 10 GeV, this has been ruled out by the CMB constraint (see discussion in Sec.~\ref{sec:constraints}). Therefore we focus on the opposite limit $\theta_1^2 e^{-2\mtn/T} < e^{-(\mtt+\mtn)/T}$, i.e., small mixing and small $\Delta$. In this case we have $\hm{C_S} > \hm{C_A} >\hm{A}$ in terms of rates. The coscattering process $\hm{S}$ has the same $\theta_1$ dependence as $\hm{C_A}$, but is less Boltzmann suppressed because $\mgd < \mtn$, so the coscattering can keep $\tt$ and $\tn$ in kinetic equilibrium after $\hm{C_A}$ freezes out. In this simple-minded picture, the DM relic density is then determined by freeze-out of $\hm{C_S}$ and $\hm{S}$. Denoting their freeze-out temperatures by $T_{\hm{C_S}}$ and $T_{\hm{S}}$, then there are two main scenarios.

\begin{figure}
\captionsetup{singlelinecheck = false, format= hang, justification=raggedright, font=footnotesize, labelsep=space}
\includegraphics[scale=0.4]{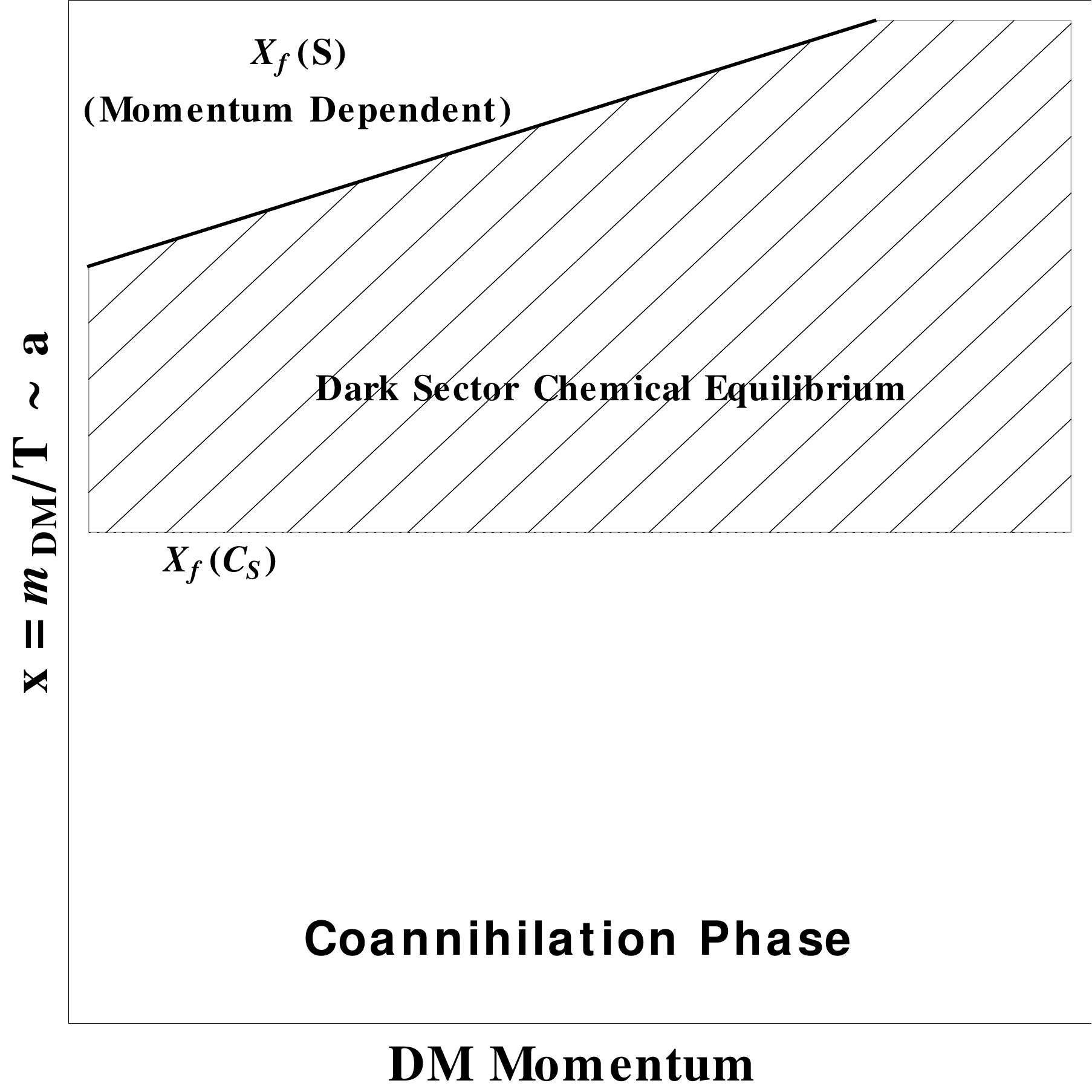}
\includegraphics[scale=0.4]{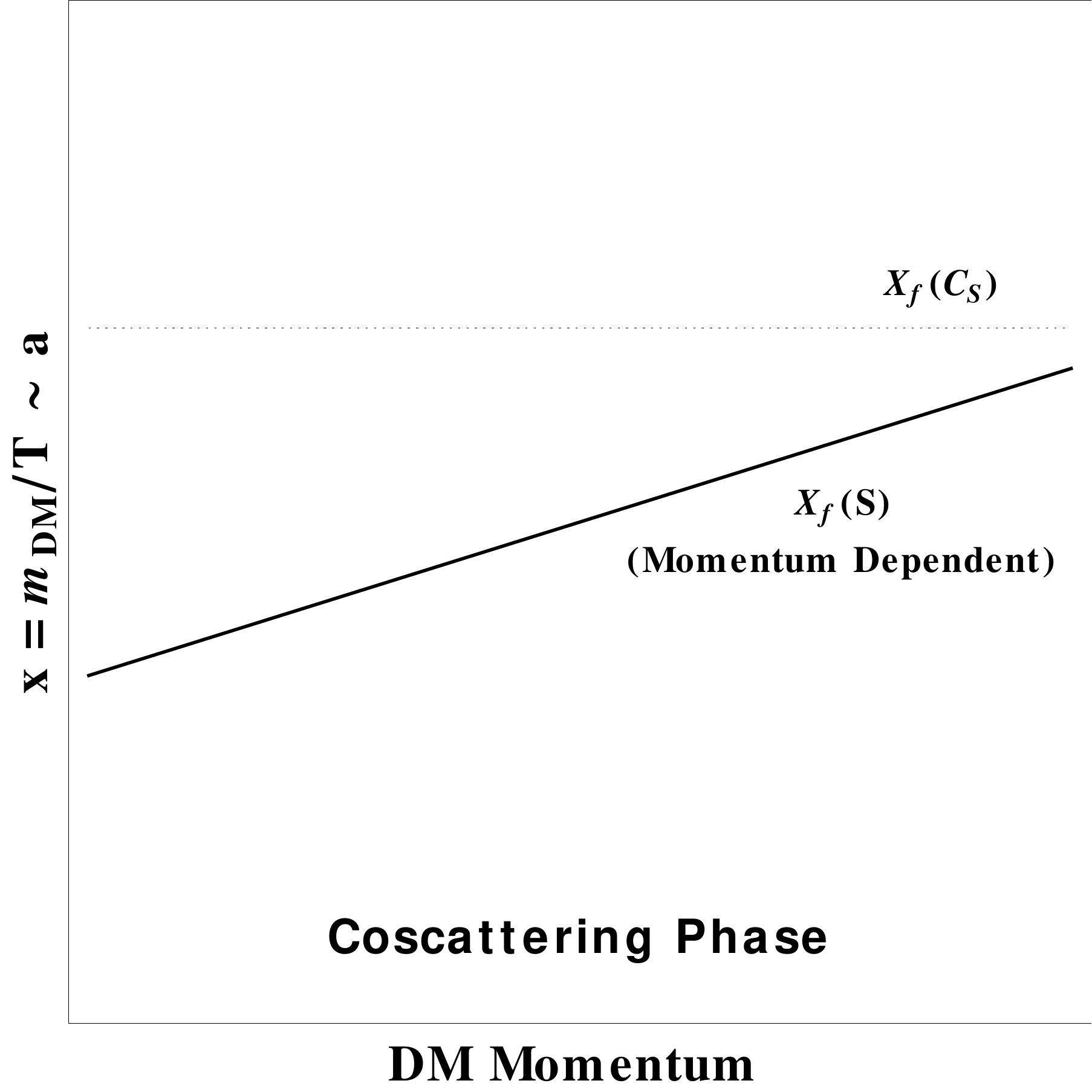}

\caption{Schematic plots of different scenarios, displayed by the freeze-out temperature and different momentum of $\tn$. {\bf Left:} Coannihilation phase where the DM relic density is dominantly determined by $\hm{C_S}$.  {\bf Right:} Coscattering phase discussed in Ref.~\cite{DAgnolo:2017dbv} where DM relic density is determined by $\hm{S}$.  }
\label{fig:Cartoons1}
\end{figure}

\begin{enumerate}
\item $T_{\hm{C_S}} > T_{\hm{S}}$: This occurs if $\theta_1^2 e^{-(\mtt+\mgd)/T} \gg e^{-2\mtt/T}$ during freeze-out so that $\hm{C_S}$ freezes out earlier. After that the total number of $\tn$ and $\tt$ in a comoving volume is fixed. The coscattering process only re-distributes the densities between $\tn$ and $\tt$, but eventually all $\tt$'s will decay down to $\tn$'s.  The DM relic density is determined by $\hm{C_S}$. This is the coannihilation phase. It is schematically depicted in the left panel of Fig.~\ref{fig:Cartoons1}. 

\item $T_{\hm{C_S}} < T_{\hm{S}}$: In the opposite limit, $\hm{S}$ freezes out before $\hm{C_S}$, and hence stops converting $\tn$ into $\tt$. On the other hand, $\hm{C_S}$ is still active and will annihilate most of the leftover $\tt$'s. The relic density in this case is determined by the coscattering process $\hm{S}$. (Remember that $\hm{A}$ has frozen out earlier.) This is the coscattering phase discovered in Ref.~\cite{DAgnolo:2017dbv}. It is illustrated in the right panel of Fig.~\ref{fig:Cartoons1}.

\end{enumerate}

\begin{figure}
\captionsetup{singlelinecheck = false, format= hang, justification=raggedright, font=footnotesize, labelsep=space}
\includegraphics[scale=0.4]{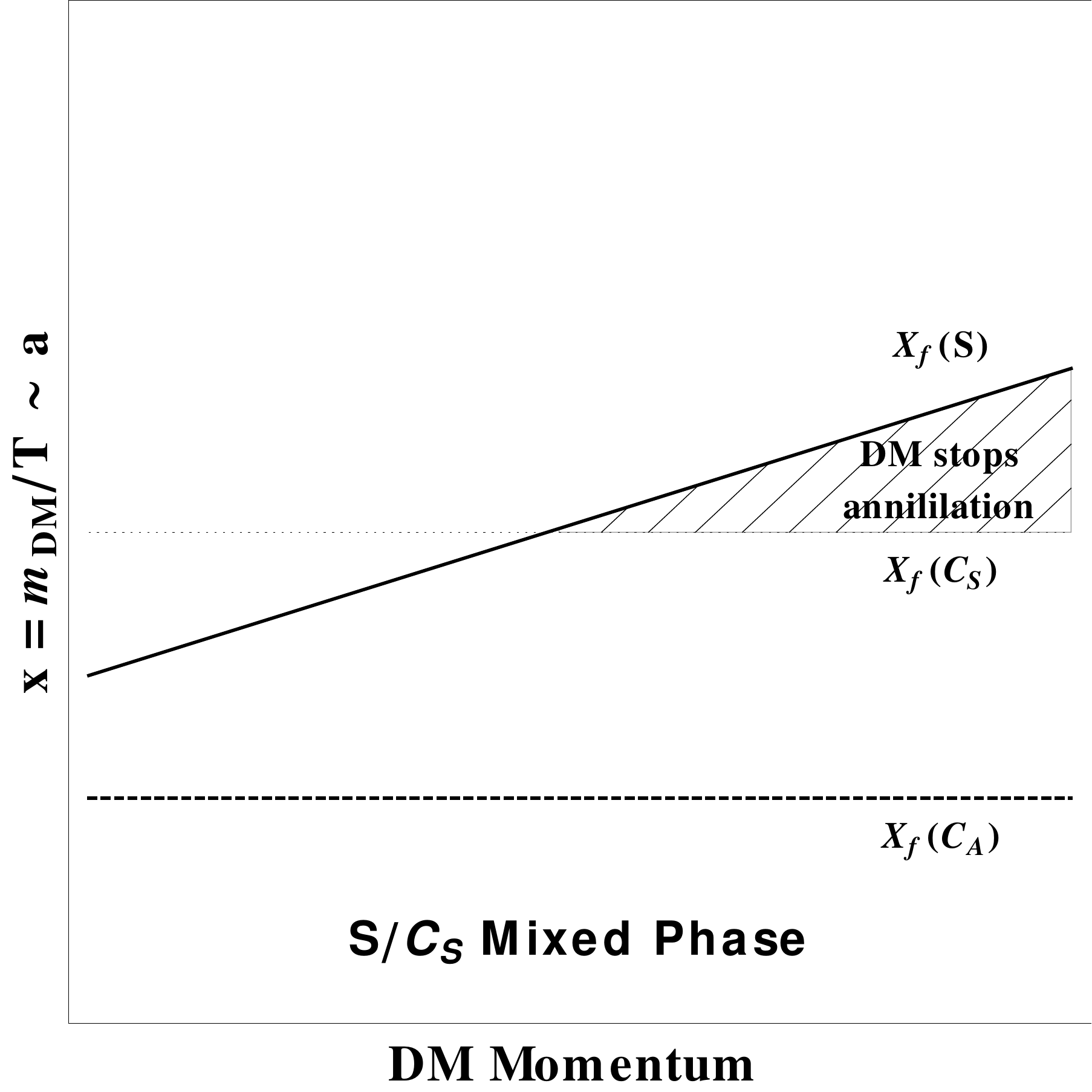}
\includegraphics[scale=0.4]{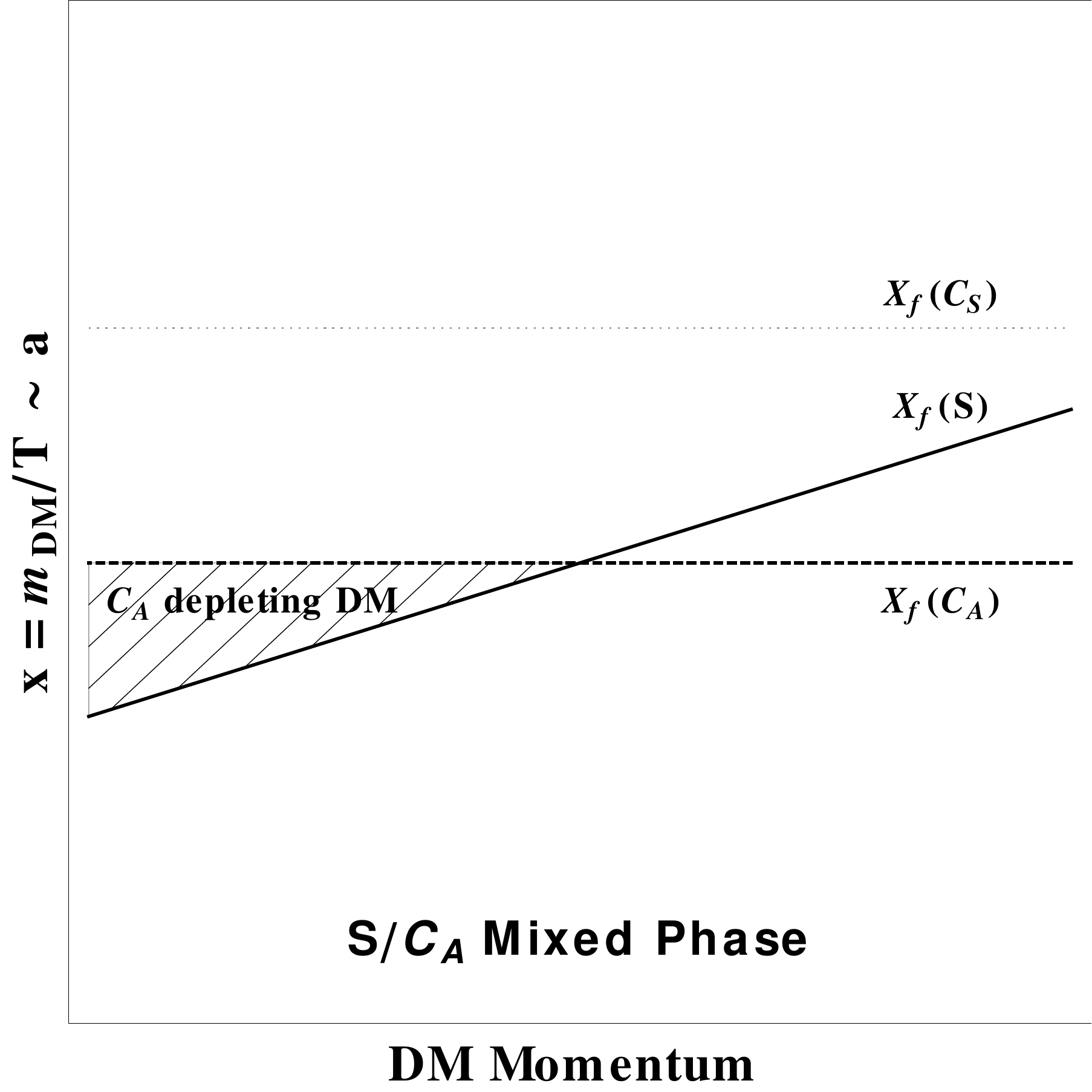}

\caption{Schematic illustrations of the mixed coscattering/coannihilation phases. {\bf Left:} The $\hm{S}/\hm{C_S}$ mixed phase: for low (high) momentum modes $\hm{S}$ freezes out earlier (later) than $\hm{C_S}$. {\bf Right:} The $\hm{C_A}/\hm{S}$ mixed phase: for low (high) momentum modes $\hm{C_A}$ freezes out earlier (later) than $\hm{S}$.}
\label{fig:Cartoons2}
\end{figure}

In the above discussion, we have associated each process with a single freeze-out temperature. This is a good approximation for the annihilation and coannihilation processes, but not for the coscattering process which has a strong momentum dependence. In the coscattering phase, the processses $\tn \gd \to \tn\gd$, $\tn\tn \to \tn\tn$ are suppressed by $\theta_1^4$ and hence are expected to freeze out earlier and cannot re-equilibrate the $\tn$ momentum. Consequently, different momentum modes in the coscattering process freeze out at different time, with low momentum modes freeze out earlier. If $\theta_1^2 e^{-(\mtt+\mgd)/T} \sim e^{-2\mtt/T}$ during freeze-out, we can have a situation that coscattering of the low momentum modes freezes out earlier than $\hm{C_S}$ while the coscattering of the high momentum modes freezes out later than $\hm{C_S}$. This is illustrated in the left panel of Fig.~\ref{fig:Cartoons2}. In this case, the relic density of low momentum modes is determined by the coscattering process and the relic density of high momentum modes is determined by the coannihilation process $\hm{C_S}$.  We have a mixed coscattering/coannihilation phase where the relic density is determined by both $\hm{S}$ and $\hm{C_S}$.

If the masses of the twin photon and the twin neutrino are close, we have $\theta_1^2 e^{-(\mtt+\mgd)/T} \sim \theta_1^2 e^{-(\mtt+\mtn)/T} <e^{-2\mtt/T}$ during freeze-out from our assumption. One expects that this belongs to the coscattering phase since $T_{\hm{C_S}} < T_{\hm{S}}$. However, the rates of $\hm{S}$ and $\hm{C_A}$ become comparable in this limit so we have $T_{\hm{S}} \sim T_{\hm{C_A}} > T_{\hm{C_S}}$. Due to the strong momentum dependence of $\hm{S}$, one can have a situation depicted in the right panel of Fig.~\ref{fig:Cartoons2}. The coscattering of low momentum modes freezes out early, but their comoving density is still reduced by the coannihilation $\hm{C_A}$ until $\hm{C_A}$ freezes out. The relic density of high momentum modes is determined by $\hm{S}$ as in the coscattering phase. In this case we have another mixed coscattering/coannihilation phase where the relic density is determined together by $\hm{C_A}$ and $\hm{S}$. Finally, if the mixing is not very small so that the rates of $\hm{C_A}$ and $\hm{C_S}$ are not far apart, the freeze-out time of the coscattering process can even cut through that of both $\hm{C_A}$ and $\hm{C_S}$, although it can only happen in some rare corner of the parameter space.  The relic density is then determined by all three processes, with the low momentum modes controlled by $\hm{C_A}$, intermediate momentum modes governed by $\hm{S}$, and high momentum modes determined by $\hm{C_S}$.

Due to the momentum dependence of the coscattering process, the relic density calculations of the coscattering and mixed phases are more complicated. We describe the calculation for each case in the next section.

\section{Relic Abundance Calculations in Various Phases}
\label{sec:calculation}

In this section we describe the calculations of DM relic abundance in different phases.

\subsection{Coannihilation}

In the coannihilation phase, the coscattering process $\hm{S}$ decouples late enough to keep $\tt$ and $\tn$ in chemical equilibrium even after all annihilation processes freeze out, so we have
\begin{equation}
\frac{n_{\tt}(T)}{n_{\tn}(T)}=\frac{n_{\tt}^{\rm{eq}}(T)}{n_{\tn}^{\rm{eq}}(T)} ,
\label{eqn:CE}
\end{equation} 
where $n$ ($n^{\rm{eq}}$) is the (equilibrium) number density. 
We can simply write down the Boltzmann equation for the total DM number density $n_{\rm{tot}}(T)=n_{\tt}(T)+n_{\tn}(T)$~\cite{Griest:1990kh,Mizuta:1992qp}:
\begin{equation}
\dot{n}_{\rm{tot}}+3Hn_{\rm{tot}}=-\left\langle\sigma v\right\rangle_{C_S}\left(n_{\tt}^2-(n_{\tt}^{\text{eq}})^2\right) 
-\left\langle\sigma v\right\rangle_{C_A}\left(n_{\tt}n_{\tn}-n_{\tt}^{\text{eq}}n_{\tn}^{\text{eq}}\right)
-\left\langle\sigma v\right\rangle_{A}\left(n_{\tn}^2-(n_{\tn}^{\text{eq}})^2\right).
\label{eqn:COA}
\end{equation}
It can be easily solved and is incorporated in the standard DM relic density calculation packages. In the parameter region that we are interested, the right-handed side is mostly dominated by the $\hm{C_S}$ term.

\subsection{Coscattering}
The DM density calculation in the coscattering phase is more involved and was discussed in detail in Ref.~\cite{DAgnolo:2017dbv}. There the authors provided an approximate solution based on the integrated Boltzmann equation:
\begin{equation}\label{eqn:boltzmannS}
\dot{n}_{\tn}+3Hn_{\tn}=-\left\langle\sigma v\right\rangle_S (n_{\tn}-n_{\tn}^{\rm eq})n_{\gd}^{\rm eq} .
\end{equation}
However, as mentioned earlier, the kinetic equilibrium of $\tn$ will not be maintained during the freeze-out of the coscattering process and different momentum modes freeze out at different time. The simple estimate from Eq.~(\ref{eqn:boltzmannS}) is not always accurate. Here we reproduce the calculation from the unintegrated Boltzmann equation to keep track of the momentum dependence.

The unintegrated Boltzmann equation of the density distribution in the momentum space $f(p, t)$ for the coscattering process $\tn(p) + \gd(k) \to \tt(p') + \gd(k')$ is given by
\begin{equation}
\left(\partial_t-H p\partial_p\right) f_{\tn}(p,t)=\frac{1}{E_p}C[f_{\tn}](p,t),
\label{eqn:collisionop}
\end{equation}
where the collision operator is defined as \cite{DAgnolo:2017dbv}
\begin{equation}
C[f_{\tn}](p,t)=\frac{1}{2}\int d\Omega_{\hm k}  d\Omega_{\hm p'} d\Omega_{\hm k'} |\overline{\mathcal{M}}|^2 [f_{\tt}(p',t)f_{\gd}(k',t)-f_{\tn}(p,t)f_{\gd}(k,t)](2\pi)^4\delta^4(\sum p^\mu).
\label{eqn:collisionop2}
\end{equation}
In the above expression $d\Omega_{\hm p}=d^3p/[(2\pi)^3 2E_p]$ is the Lorentz-invariant integration measure and $|\overline{\mathcal{M}}|^2$ is the squared amplitude averaged over initial and summed over final state quantum numbers.

In this phase $\tt$ and $\gd$ can be assumed to be in thermal equilibrium with the thermal bath, $f_{\tt(\gd)} = f^{\rm{eq}}_{\tt(\gd)}$, from the processes $\tt \gd \leftrightarrow \tt \gd$, $\tt \tt \leftrightarrow \gd \gd$ and $\gd$ interactions with SM fields. Using $f^{\rm{eq}}_{\tt} (p', t) f^{\rm{eq}}_{\gd} (k', t) = f^{\rm{eq}}_{\tn}(p, t) f^{\rm{eq}}_{\gd} (k,t)$, 
 the right-hand side of Eq.~(\ref{eqn:collisionop}) can be simplifed as
\begin{equation}\label{eqn:c3}
\frac{1}{E_p}C[f_{\tn}](p,t)=[f_{\tn}^{\text{eq}}(p,t)-f_{\tn}(p,t)]\widetilde{C}(p,t),
\end{equation}
where the reduced collision operator $\widetilde{C}(p,t)$ takes the form,
\begin{align}
\widetilde{C}(p,t) &=\frac{1}{2E_p}\int d\Omega_{\hm k} f_{\gd}^{\text{eq}}(k,t)\int d\Omega_{\hm p'} d\Omega_{\hm k'}|\overline{\mathcal{M}}|^2 (2\pi)^4\delta^4({p^\mu}+{k^\mu}-{p^{\prime\mu}}-{k^{\prime\mu}})\\
&=\frac{1}{2E_p}\int d\Omega_{\hm k} f_{\gd}^{\text{eq}}(k,t) j(s)\sigma(s),
\label{eqn:coredefinition}
\end{align}
with the Lorentz-invariant flux factor $j(s) = 2E_p 2 E_k |\hm{v}_p - \hm{v}_k| $. The calculation of $\widetilde{C}$ from the coscattering is described in Appendix~\ref{sec:collision}.

The left-hand side of Eq.~(\ref{eqn:collisionop}) can be written as a single term by defining the comoving momentum $q\equiv p \,a$,
then Eq.~(\ref{eqn:collisionop}) becomes a first order differential equation of  
the scale factor $a$ for each comoving momentum $q$:
\begin{equation}
Ha\partial_a f_{\tn}(q,a) =[f_{\tn}^{\text{eq}}(q,a)-f_{\tn}(q,a)]\widetilde{C}(q,a),
\label{eqn:collisioncore}
\end{equation}
where we have written the distribution $f$ as a function of $q$ and $a$, instead of $p$ and $t$.
Taking the boundary condition $f_{\tn} (q, a_0) = f^{\rm eq}_{\tn} (q, a_0)$ at an early time $a_0$, the solution is given by 
\begin{equation}
f_{\tn}(q,a)=f_{\tn}^{\text{eq}}(q,a)-\int_{a_0}^a da' \frac{d f_{\tn}^{\text{eq}}(q,a')}{d a'} e^{-\int_{a'}^a \frac{\widetilde{C}(q,a^{\prime\prime})}{ H a^{\prime\prime}}da^{\prime\prime}}.
\label{eqn:coscattersolution}
\end{equation}
At early time and high temperature where $\widetilde{C}(q,a) \gg H(a)$, the second term on the right-hand side can be neglected and $\tn$ density is given by the equilibrium density as expected. At late time and low temperature, $\widetilde{C}(q,a) \ll H(a)$, the $\tn$ comoving density stops changing. For each comoving momentum $q$, one can find a time $a' = a_f(q)$ beyond which the exponent is small so that the exponential factor is approximately 1. Then the final $\tn$ density is roughly given by $f^{\rm eq}_{\tn}( q, a_f(q))$.  The $a_f(q)$ can be viewed as the freeze-out scale factor for coscattering of the comoving momentum $q$ of $\tn$. It is roughly determined by  $\widetilde{C}(q,a_f(q))\simeq H(a_f(q))$.

As mentioned in the previous section, if $\mtt > \mtn +\mgd \, (r<\Delta)$, the inverse decay $\tn + \gd \to \tt$ plays a similar role as the coscattering process. Its contribution should be added to the collision operator, which is calculated in Appendix~\ref{sec:ID}. It is larger than the coscattering contribution, as it requires less energy to produce the final state. In the parameter region that we consider, it always makes the second term in Eq.~(\ref{eqn:coscattersolution}) negligible before $\hm{C_S}$ freezes out. Therefore, it goes back to the coannihilation phase if this on-shell decay is allowed.

\subsection{Mixed Phases}

From the discussion in the previous section, mixed phases occur when $T_{\hm{S}} \sim T_{\hm{C_S}}$ or $T_{\hm{S}} \sim T_{\hm{C_A}}$. We first consider the  case 
$T_{\hm{ C_A}}\sim T_{\hm S} > T_{\hm{C_S}}$ ($\hm{C_A}/\hm{S}$ mixed phase). This happens when $\mgd \simeq \mtn$. Because the coannihilation process $\hm{C_A}$, $\tn(p) + \tt(k) \to \gd(p') + \gd(k')$, is also important in this case, the contribution from $\hm{C_A}$ to the collision operator, 
\begin{equation}
\widetilde{C}_{\hm{C_A}}(p,t) =\frac{1}{2E_p}\int d\Omega_{\hm k} f_{\tt}^{\text{eq}}(k,t)\int d\Omega_{\hm p'} d\Omega_{\hm k'}|\overline{\mathcal{M}}_{\hm{C_A}}|^2 (2\pi)^4\delta^4({p}+{k}-{p'}-{k'}),
\end{equation}
should be included in Eq.~(\ref{eqn:collisionop})
in addition to the coscattering contribution of Eq.~(\ref{eqn:collisionop2}).
Since $T_{\hm{C_S}}$ is assumed to be small, $\tt$ stays in thermal equilibrium during the decoupling of $\hm{ S}$ and $\hm{ C_A}$. 
In contrast to $\hm S$, there is no kinematic threshold in $\hm{ C_A}$, so it has very weak momentum dependence. Hence we can treat it as a function of time/temperature only. In Eq.~(\ref{eqn:coscattersolution}) we can replace $\widetilde{C}$ by $\widetilde{C}_{\hm S}+\widetilde{C}_{\hm{C_A}}$ and conduct the computation in the same manner as in the coscattering phase. In this $\hm{C_A}/\hm{S}$ mixed phase, $\widetilde{C}_{\hm{C_A}}> \widetilde{C}_{\hm S}$ for low momentum modes so their freeze out temperature is determined by $\hm{C_A}$, while for high momentum modes $\widetilde{C}_{\hm S}> \widetilde{C}_{\hm{C_A}}$ and their contributions is given by the coscattering result, as illustrated in the right panel of Fig.~\ref{fig:Cartoons2}

For the other mixed phase ($\hm{S}/\hm{C_S}$) where $T_{\hm{ C_S}}\sim T_{\hm{ S}}< T_{\hm{C_A}}$, the calculation is more complicated.
In this case $\tt$ is no longer in thermal equilibrium with the thermal bath, $f_{\tt}$ itself is unknown, hence cannot be set to equal $f_{\tt}^{\rm eq}$. As a result, Eq.~(\ref{eqn:c3}) no longer holds. Moreover, different from the usual coannihilation scenario, $\tt$ and $\tn$ are not in chemical equilibrium. A complete solution requires solving the two coupled Boltzmann equations for $\tt$ and $\tn$ in this case, which is numerically expensive. However, we can assume that $\tt$ is still in kinetic equilibrium with the SM sector (i.e., has the canonical distribution up to an unknown overall factor) due to the elastic scattering with $\gd$. We can write 
$f_{\tt}/f_{\tt}^{\rm eq}=Y_{\tt}/Y_{\tt}^{\rm eq}$ where $Y_{\tt}= n_{\tt}/s$ is the comoving number density. The term in the square bracket of Eq.~(\ref{eqn:collisionop2}) can be written as:
\begin{eqnarray}
& &f_{\tt}(p',t)f_{\gd}(k',t)-f_{\tn}(p,t)f_{\gd}(k,t) = f^{\text{eq}}_{\tt}(p',t)\frac{Y_{\tt}(t)}{Y_{\tt}^{\text{eq}}(t)}f_{\gd}^{\text{eq}}(k',t)-f_{\tn}(p,t)f_{\gd}^{\text{eq}}(k,t)\\ 
&=&
f_{\gd}^{\text{eq}}(k, t)[\frac{Y_{\tt}(t)}{Y_{\tt}^{\text{eq}}(t)}f^{\text{eq}}_{\tn}(p,t)-f_{\tn}(p,t)].\nonumber
\label{eqn:tauratio}
\end{eqnarray}
This corresponds to replacing $f_{\tn}^{\rm eq}(q, a)$ by $(Y_{\tt}(a)/ Y^{\rm eq}_{\tt} (a)) f_{\tn}^{\rm eq}(q, a)$ in  Eq.~(\ref{eqn:collisioncore}).  The solution Eq.~(\ref{eqn:coscattersolution}) is modified to~\cite{Garny:2017rxs}
\begin{equation}
f_{\tn}(q,a)=\frac{Y_{\tt}(a)}{Y_{\tt}^{\text{eq}}(a)}f_{\tn}^{\text{eq}}(q,a)+\int_{a_0}^a da' \frac{-d \frac{Y_{\tt}(a')}{Y_{\tt}^{\text{eq}}(a')}f_{\tn}^{\text{eq}}(q,a')}{d a'} e^{-\int_{a'}^a \frac{\widetilde{C}(q,a'')}{ H a''}da''}.
\label{eqn:intersolution}
\end{equation}

Of course we do not know $Y_{\tt}(a)$ in advance. One way to solve this problem is to employ the iterative method~\cite{Garny:2017rxs} which is described in Appendix~\ref{sec:iteration}. One starts with some initial guess of $Y_{\tt} (a)$ to obtain the solution for Eq.~(\ref{eqn:intersolution}), then use that solution in the Boltzmann equation for $n_{\tt}$ to obtain a new $Y_{\tt}(a)$, and repeat the procedure until the result converges. We find that a good first guess is to simply take $Y_{\tt}(a)$ to be the one obtained in the coannihilation calculation. In coannihilation, $\tt$ and $\tn$ are in chemical equilibrium:
\begin{equation}
\frac{Y_{\tt}(a)}{Y_{\tt}^{\text{eq}}(a)}f_{\tn}^{\text{eq}}(q,a)\simeq \frac{Y_{\tn}(a)}{Y_{\tn}^{\text{eq}}(a)}f_{\tn}^{\text{eq}}(q,a) = f^{\text{CA}}_{\tn}(q,a)\simeq f^{\text{CA}}_{\text{tot}}(q,a),
\end{equation} 
where the superscript CA indicates that the result is obtained from the coanihillation-only estimation (Eq.~(\ref{eqn:COA})).
Notice that in this case the first term in Eq.~(\ref{eqn:intersolution}) is simply the contribution from coannihilation. If $T_{\hm{S}} < T_{\hm{C_S}}$, $\hm{S}$ will still be active when $\hm{C_S}$ freezes out, $\widetilde{C}(q, a'')/(Ha'')$ will be large and the second term in Eq.~(\ref{eqn:intersolution}) will be suppressed. We obtain the correct coannihilation limit. On the other hand, if $T_{\hm{S}} >T_{\hm{C_S}}$, the coannihilation $\hm{C_S}$ is effective to keep ${Y_{\tt}(a)}/{Y_{\tt}^{\text{eq}}(a)}\approx1$ and Eq.~(\ref{eqn:intersolution}) returns to the coscattering result in Eq.~(\ref{eqn:coscattersolution}). Using $Y_{\tt}(a)$ from the coannihalation calculation in  Eq.~(\ref{eqn:intersolution}) gives the correct results in both the coannihilation and coscattering limits. The expression interpolates between these two limits in the mixed phase and it turns out to be an excellent approximation to the correct relic density even without performing the iterations. (See Appendix~\ref{sec:iteration}.)

\begin{figure}
\captionsetup{singlelinecheck = false, format= hang, justification=raggedright, font=footnotesize, labelsep=space}
\includegraphics[scale=0.28]{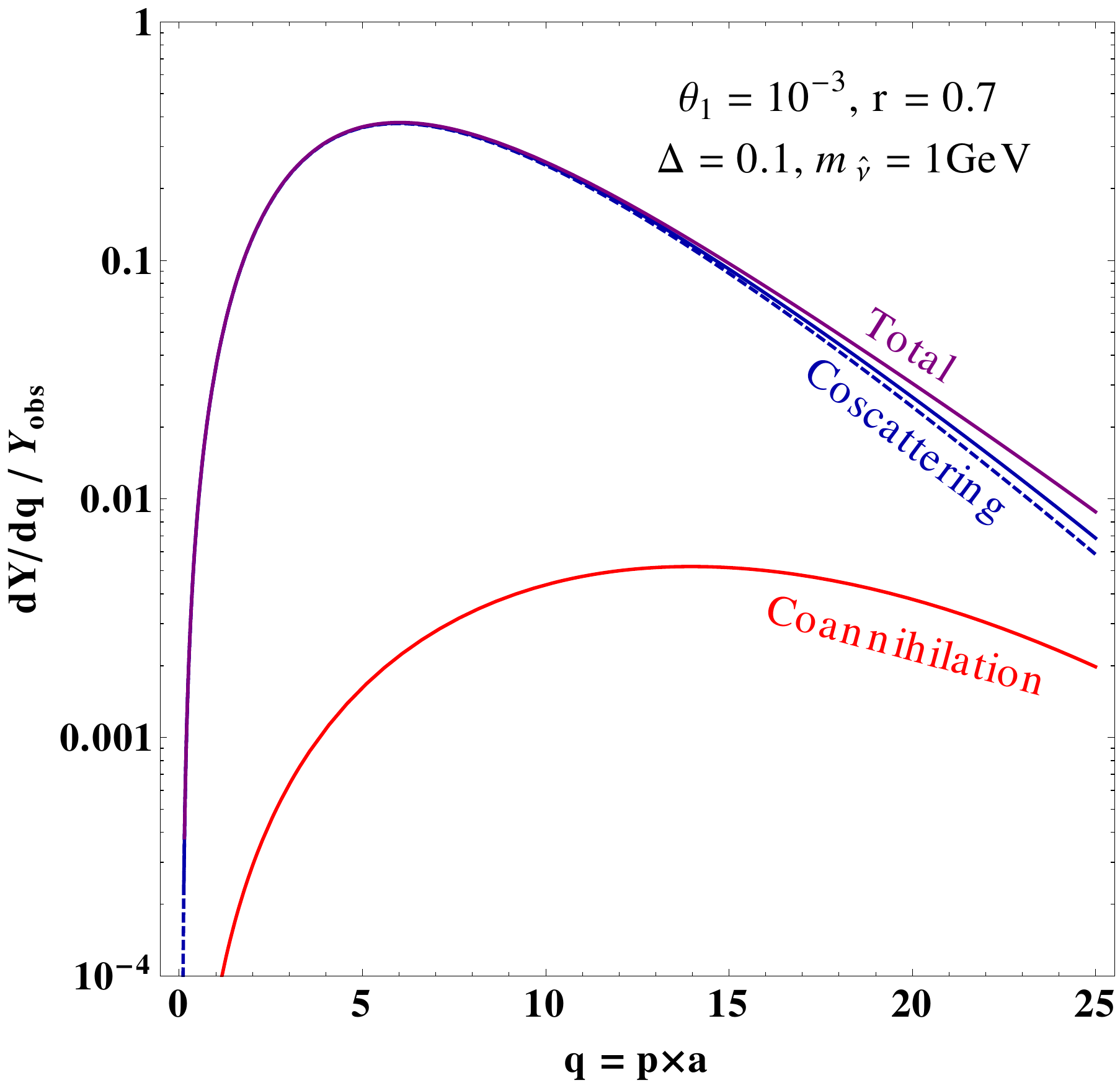}
\includegraphics[scale=0.28]{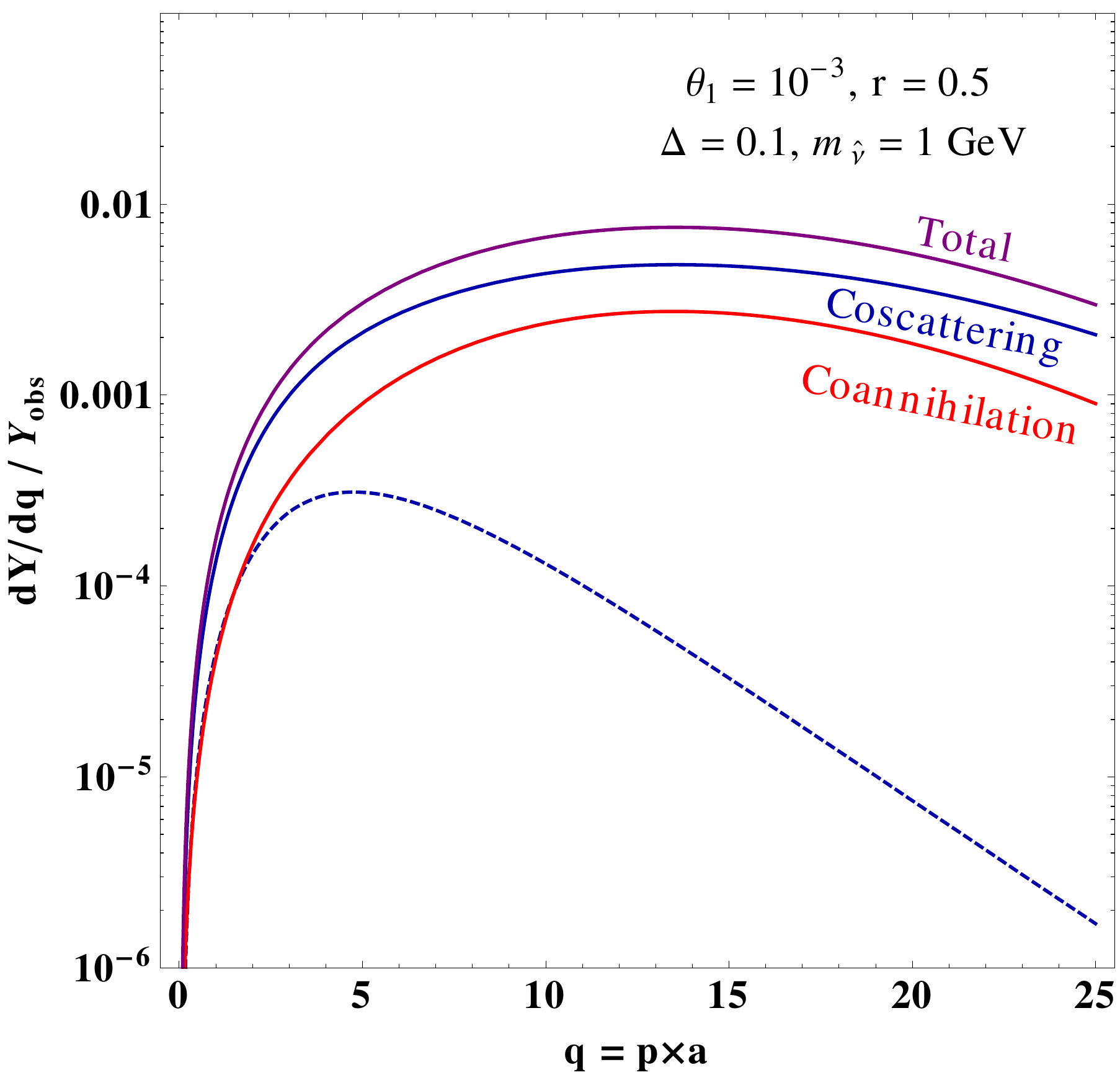}
\includegraphics[scale=0.28]{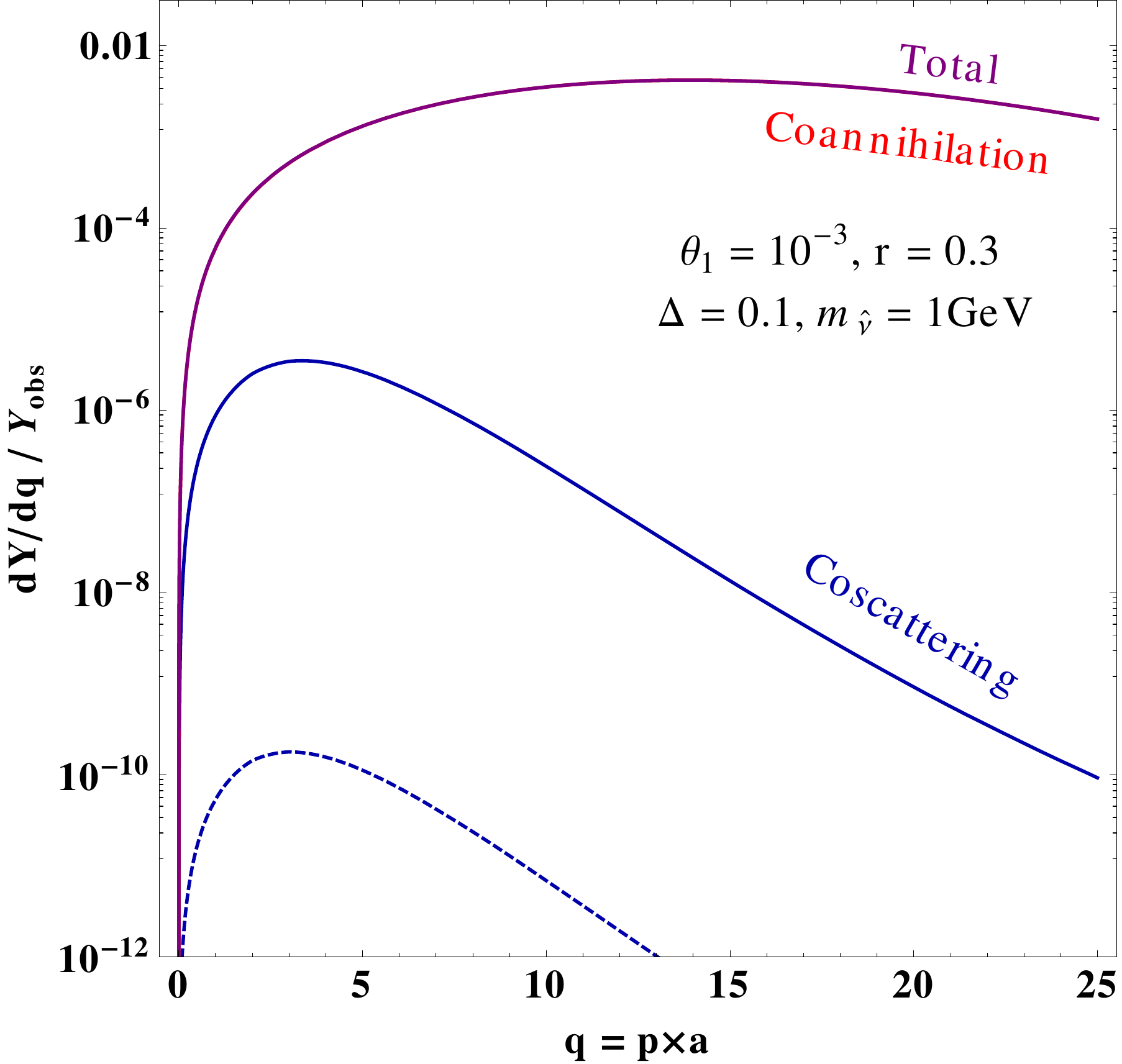}
\caption{Benchmark results for relic density calculation in the coscattering (left), mixed (central), and coannihilation (right) phases. The red curves are the contributions from the first term of Eq.~(\ref{eqn:intersolution}) which corresponds to a pure coannihilation calculation. The blue solid curves are due to the coscattering contributions from the second term of Eq.~(\ref{eqn:intersolution}). The purple curves are the total contributions. We can see that in the coscattering phase and the coannihilation phase the total contribution is dominated by one term, while in the mixed phase both terms give comparable contributions. The blue dashed curves are calculated from the coscattering formula of Eq.~(\ref{eqn:collisionop}). Only in the coscattering phase the blue dashed curve approximates the correct result. }
\label{fig:YratioBM123}
\end{figure}
For completeness, we can include the $\hm{C_A}$ contribution in Eq.~(\ref{eqn:intersolution}), then this result also applies to the case if $T_{\hm{S}}$ cuts through both $T_{\hm{C_A}}$ and $T_{\hm{C_S}}$.  Fig.~\ref{fig:YratioBM123} shows the differential DM density as a function of the comoving momentum in different phases, all calculated from Eq.~(\ref{eqn:intersolution}) including all contributions. We will use this formula for numerical calculations of DM densities in the next section in all phases.

\section{Numerical results }
\label{sec:numerical}

In this section we present the numerical calculations of the dark matter relic abundance for various parameter choices of the model. We fix $\hat{e} =0.3$ for the twin electromagnetic gauge coupling and calculate the DM density in units of the observed $\Omega_{\text{obs}} h^2 = 0.12$. All calculations are performed using Eq.~(\ref{eqn:intersolution}) including contributions from all relevant processes. 
\begin{figure}
\centering
\captionsetup{singlelinecheck = false, format= hang, justification=raggedright, font=footnotesize, labelsep=space}
\includegraphics[scale=0.4]{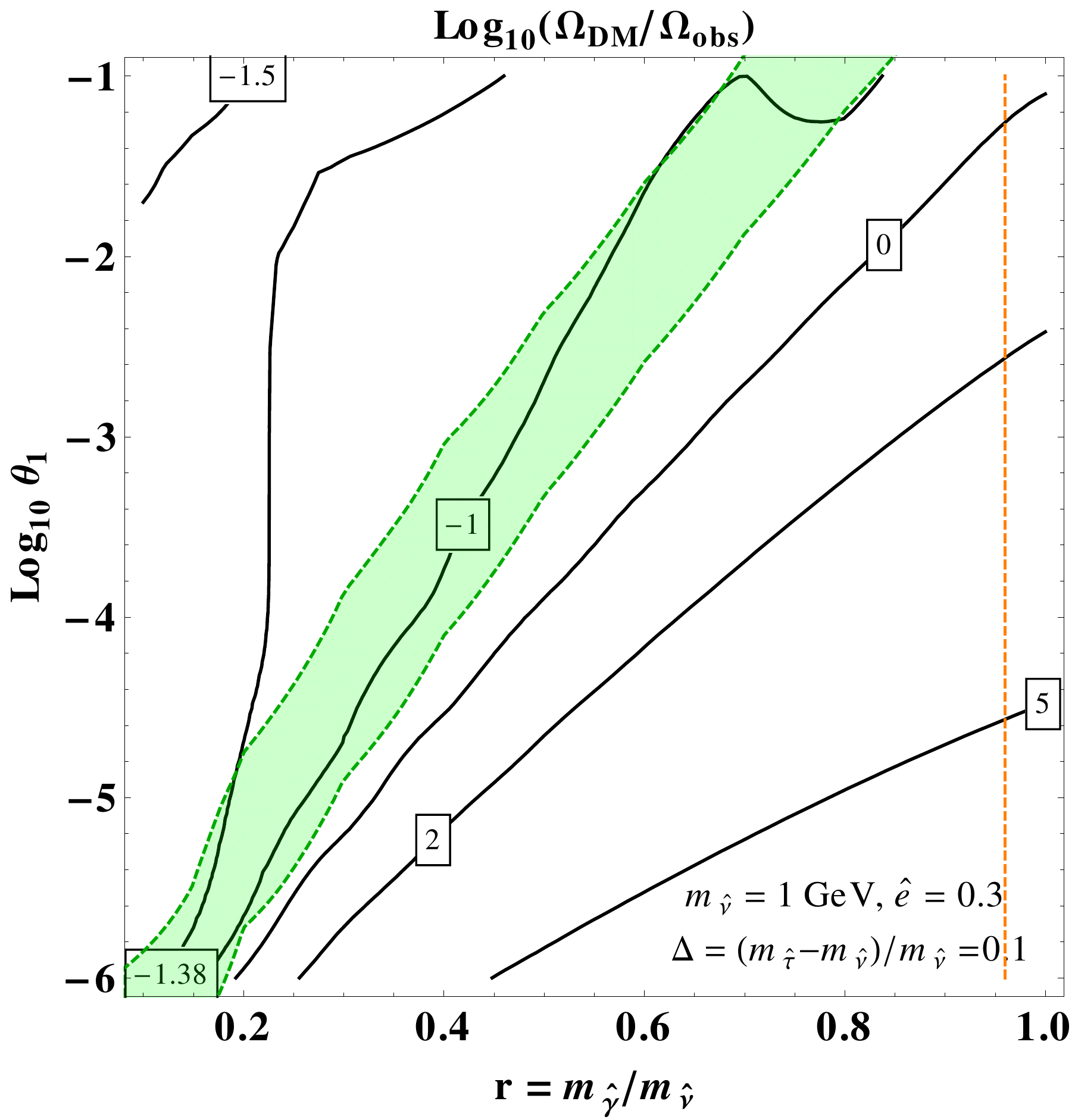}
\includegraphics[scale=0.4]{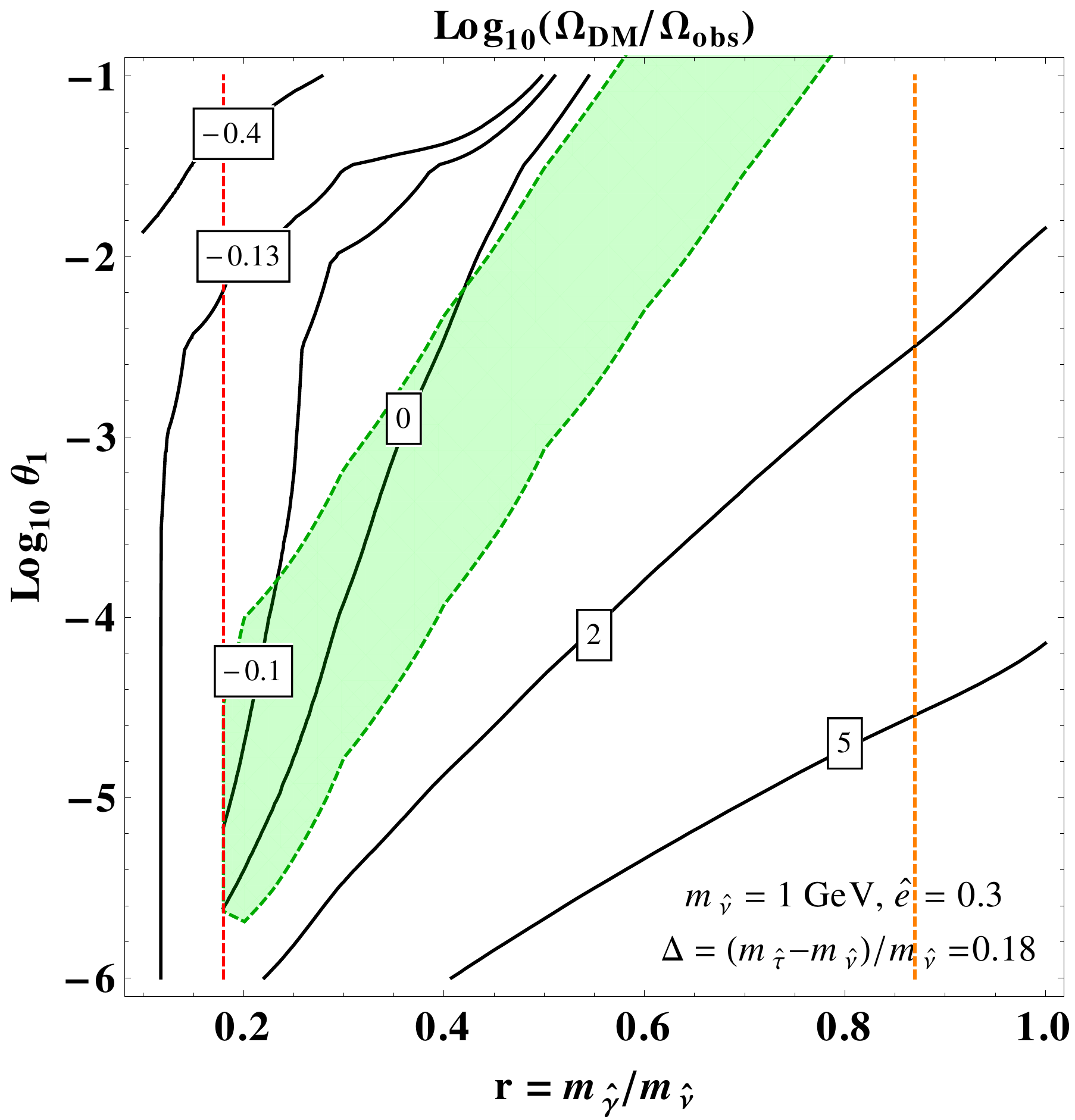}
\caption{Relic density for $\mtn=1$~GeV and $\Delta = 0.1$ and 0.18. Contours of fixed $\log_{10}(\Omega/\Omega_{\rm{OBS}})$ values are depicted. The shaded green region indicates the mixed phase where $T_{\hm S}\sim T_{\hm{C_S}}$. The region to the lower right of the green band is in the  coscattering phase while the coannihilation phase is in the upper left corner. In the region to the right of the orange dashed line, $\hm{C_A}$ becomes important and it enters the $\hm{C_A}/\hm{S}$ mixed phase. The contours become slightly less sensitive to $\mgd$ or $r$.  The red dashed line indicates $r=\Delta$. To its left the $\tt \leftrightarrow \gd+\tn$ decay and inverse decay are open. Their large rates keep $\tt,\tn$ in chemical equilibrium, making this region coannihilation-like. 
}
\label{fig:Omega_1G}
\end{figure}

In Fig.~\ref{fig:Omega_1G}, we consider $\mtn= 1$~GeV and plot the DM density dependence on $r =\mgd/\mtn$ and the mixing angle $\theta_1$, for $\Delta = (\mtt-\mtn)/\mtn = 0.1$ and $0.18$. The contours are in $\log_{10} (\Omega_{\rm DM}/ \Omega_{\rm obs})$ and the ``0'' contour represents points which produce the observed DM density. 
The coscattering phase sits in the lower-right region, as for small $\theta_1$ and large $r$ the coscattering process $\hm{S}$ is suppressed and freezes out earlier. The upper-left region, on the other hand, belongs to the coannihilation phase. The green band separating them corresponds to the $\hm{S}/\hm{C_S}$ mixed phase. The boundaries of the green band are determined by the condition $T_{\hm{S}} (q=p\,a=0) = T_{\hm{C_S}}$ and $T_{\hm{S}} (q=25) = T_{\hm{C_S}}$. The contribution to the relic density from modes with $q>25$ is small and is ignored in the coscattering calculation. The orange vertical dashed line corresponds to $T_{\hm{S}} (q=0) = T_{\hm{C_A}}$. To the right of it $\hm{C_A}$ becomes relevant and we enter the $\hm{C_A}/ \hm{S}$ mixed phase. The DM relic density is slightly reduced by $\hm{C_A}$ compared to a pure coscattering calculation. The red vertical dashed line indicates $r=\Delta$. To the left of it the on-shell decay and inverse decay $\tt \leftrightarrow \tn \gd$ are open so this whole region is in the coannihilation phase. 

The relic density in the coannhilation phase is mostly independent of $\theta_1$ because it is mainly controlled by $\hm{C_S}$ which hardly depends on $\theta_1$. Only at larger $\theta_1$ values when $\hm{C_A}$ and $\hm{A}$ become relevant the DM relic density shows some $\theta_1$ dependence. The dependence on $r$ of the relic density in the coannihilation phase is also mild, as it mainly affects the phase space of the coannihilation process. In the coscattering phase, the DM relic density increases as $\Delta$ increases, which can be seen by comparing the two plots in Fig.~\ref{fig:Omega_1G}. This is because a larger gap between $\mtt$ and $\mtn$ requires a higher threshold momentum for $\gd$ to make $\hm{S}$ happen. Therefore the coscattering is suppressed, resulting in a larger relic density.  For $\mtn= 1$~GeV, $\Delta=0.1$, the observed DM density is produced in the coscattering phase, while for $\Delta = 0.18$ it moves to the mixed phase or coannihilation phase.

\begin{figure}[ht]
\centering
\captionsetup{singlelinecheck = false, format= hang, justification=raggedright, font=footnotesize, labelsep=space}
\includegraphics[scale=0.4]{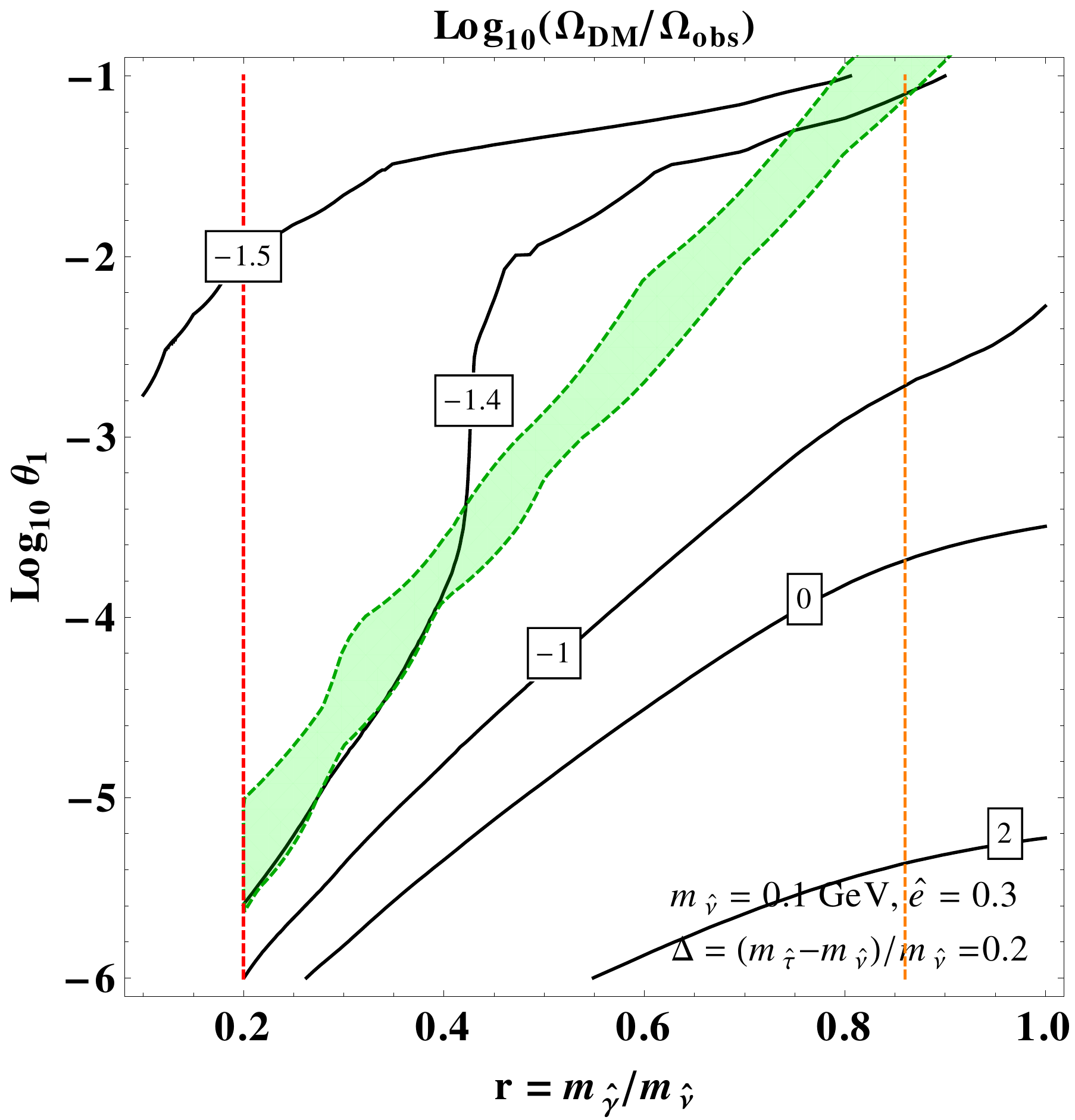}
\includegraphics[scale=0.4]{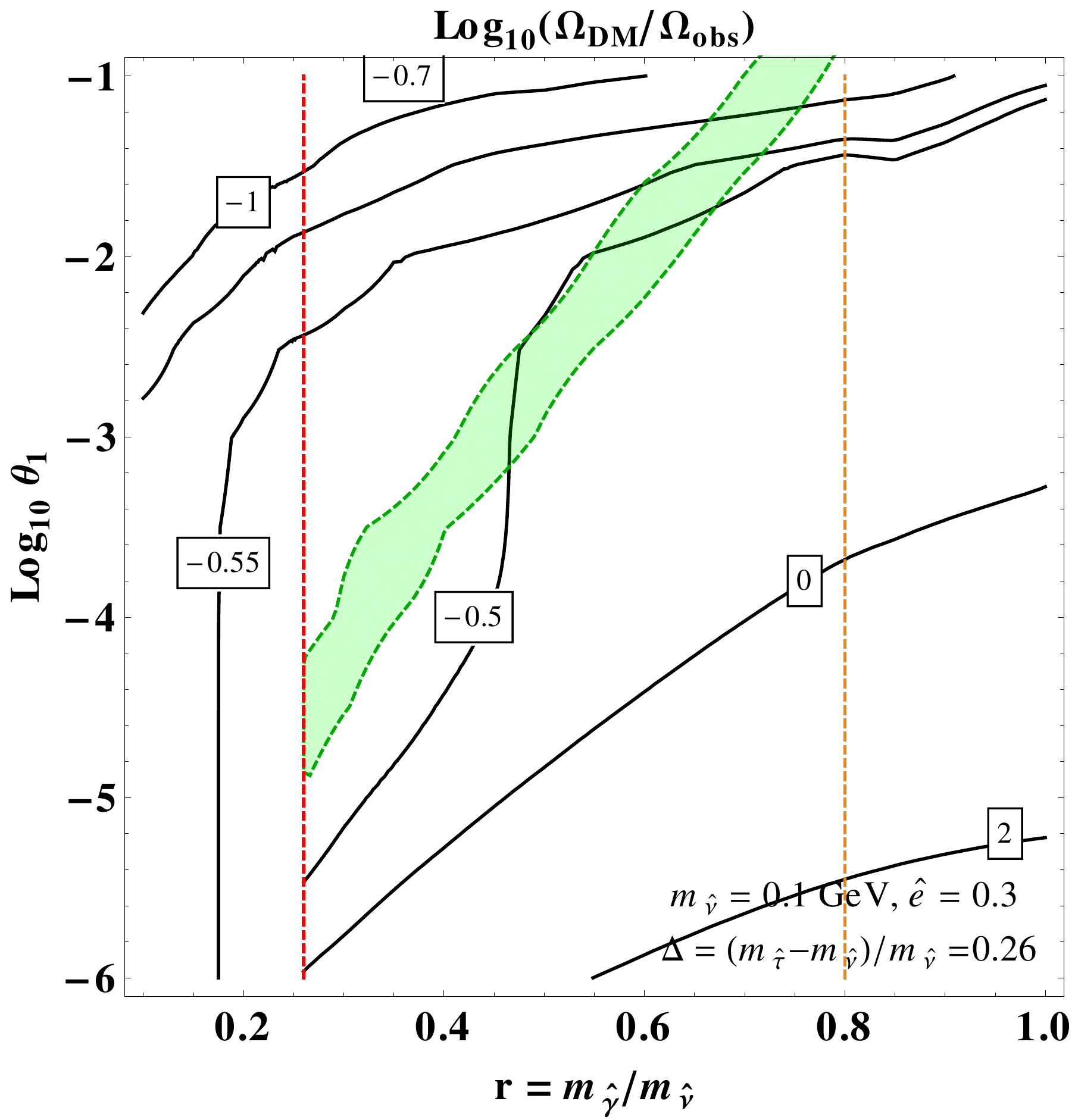}
\caption{Similar to Fig.~\ref{fig:Omega_1G}, but for $\mtn=100$~MeV and $\Delta = 0.2$ and 0.26.}
\label{fig:Omega_100M}
\end{figure}

\begin{figure}[ht]
\centering
\captionsetup{singlelinecheck = false, format= hang, justification=raggedright, font=footnotesize, labelsep=space}
\includegraphics[scale=0.4]{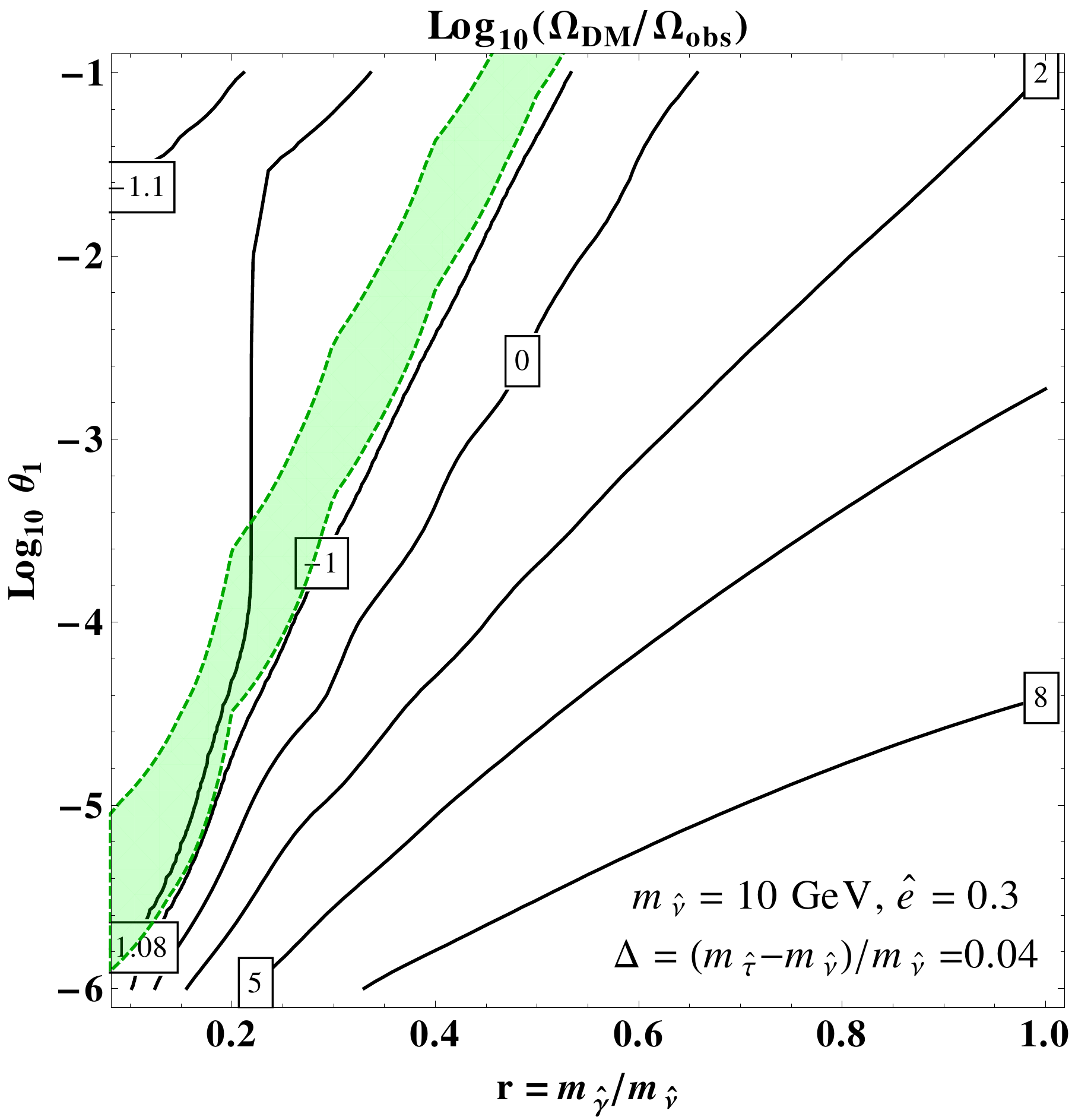}
\includegraphics[scale=0.4]{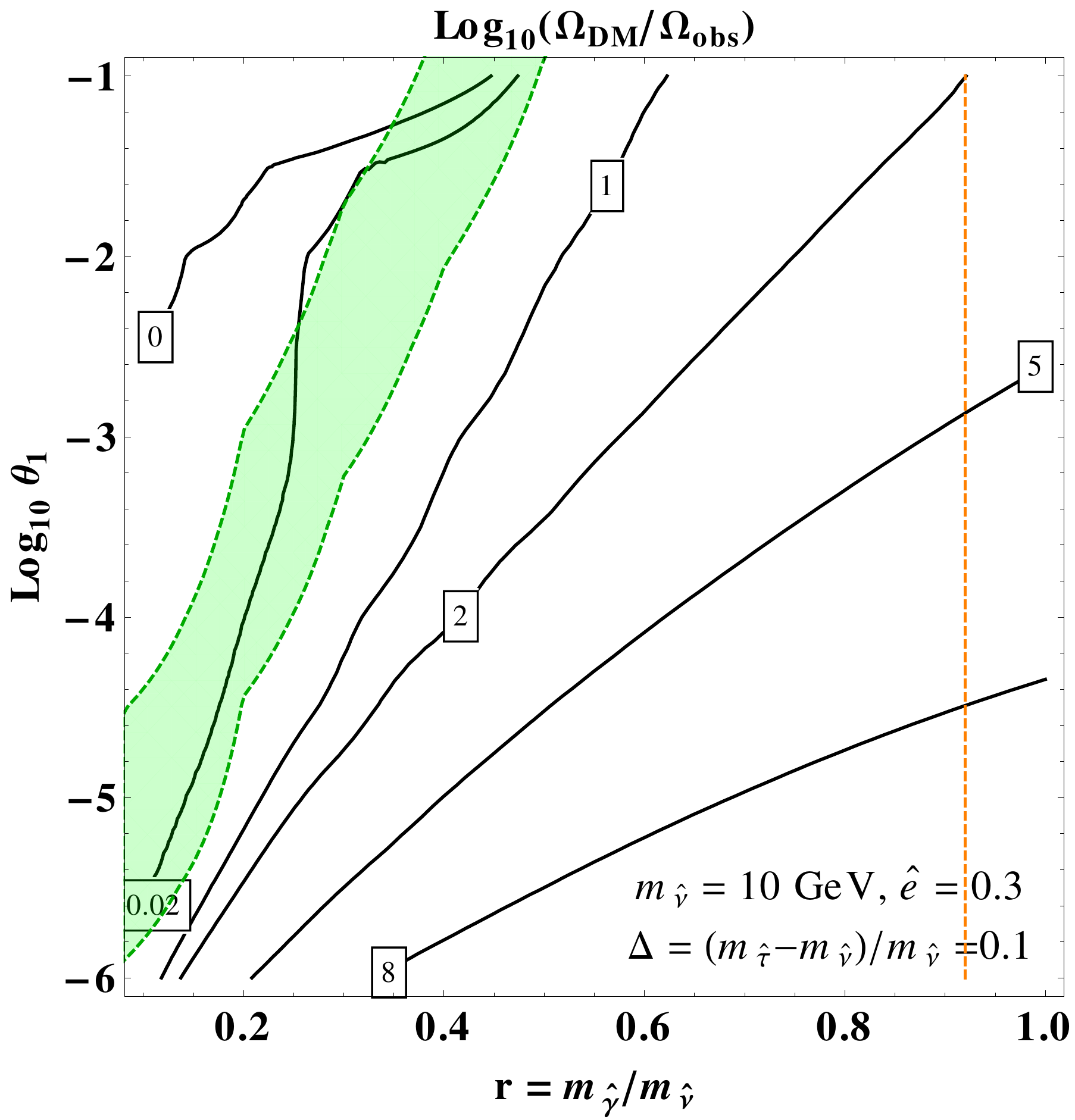}
\caption{Similar to Fig.~\ref{fig:Omega_1G}, but for $\mtn=10$~GeV and $\Delta = 0.04$ and 0.1.}
\label{fig:Omega_10G}
\end{figure}

In Figs.~\ref{fig:Omega_100M} and \ref{fig:Omega_10G} we show the results for $\mtn=100$~MeV and 10 GeV. A larger DM mass will give a larger relic density if all other parameters are fixed. Consequently for a correct relic density we need a smaller (larger) $\Delta$ for a larger (smaller) $\mtn$. The $\Delta$ values are chosen to be 0.2 and 0.26  for the two plots with $\mtn=100$~MeV, and 0.04 and 0.1 for the two plots with $\mtn= 10$~GeV. The behaviors of the contours are similar to the case of $\mtn=1$~GeV.

\section{Experimental Constraints and Tests}
\label{sec:constraints}

In this section we discuss the current experimental constraints and future experimental tests of this model. A comprehensive summary for this type of DM scenarios can be found in Ref.~\cite{DAgnolo:2018wcn}. 

\subsection{Direct Detection} 

The dark matter $\tn$ interacts with SM particles through the Higgs portal or dark photon portal. The Higgs portal is suppressed by the Higgs mixing between the SM and twin sectors, and the twin neutrino Yukawa coupling. The dark photon portal is suppressed by the kinetic mixing $\epsilon$, and also the mixing angle $\theta_1$. Most current DM direct detection experiments are based on heavy nuclei recoiling when the nuclei scatter with the DM particles. They become ineffective for light DM less than a few GeV, but could potentially constrain the parameter space of heavier twin neutrino region, as the Yukawa coupling also becomes larger for a heavier twin neutrino. For $\mtn$=10~GeV and $f/v =3$, the $\tn$-nucleon cross section is of $\mathcal{O}(10^{-47})$~cm$^2$,  which is dominated by the Higgs portal. Such $\tn$ DM is not yet constrained by recent liquid xenon DM detectors~\cite{daSilva:2017swg,Aprile:2017iyp,Cui:2017nnn}. A conservative estimate according to Ref.~\cite{Djouadi:2011aa} gives an upper limit of $\sim$22(60)~GeV on $\mtn$ for $f/v =3(5)$.

For even lighter DM, recent upgrades/proposals of detecting DM-electron scattering can largely increase the sensitivity for sub-GeV mass DM~\cite{Essig:2013lka,Alexander:2016aln}. In our model, the $\tn$-$e$ couplings from both the dark photon portal and the Higgs portal are highly suppressed. The elastic cross section of $\tn-e$ scattering from the dark photon portal is
\begin{eqnarray}
\sigma_{\nu e} &\simeq& \frac{g^2 \y^2 \epsilon^2 \theta_1^4}{\pi \mgd^4}\bigg(\frac{m_e \mgd}{m_e+\mgd}\bigg)^2 \nonumber \\
&\simeq& 4.3\times10^{-38} \bigg(\frac{\y}{0.3}\bigg)^2 \bigg(\frac{\epsilon}{10^{-3}}\bigg)^2\bigg(\frac{\theta_1}{10^{-1}}\bigg)^4 \bigg(\frac{10~\text{MeV}}{\mtn}\bigg)^4 \bigg(\frac{0.5}{r}\bigg)^4  ~~\text{cm}^2,
\end{eqnarray}
where the reference values of mixing parameters $\epsilon$, $\theta_1$ have been chosen close to the upper bounds to maximize the cross section. This is not yet constrained by recent electron-scattering experiments, including SENSEI~\cite{Crisler:2018gci}, Xenon10~\cite{Essig:2012yx}, DarkSide-50~\cite{Agnes:2018oej}. Future upgrades will be able to probe part of the parameter space with large mixings.

It is also worth mentioning that there are also crystal experiments based on phonon signals coming from DM scattering off nuclei in the detector, such as CRESST-III~\cite{Petricca:2017zdp}. Thanks to the low energy threshold ($\mathcal{O}$(50~eV)), these experiments will also be sensitive to the sub-GeV DM mass region. For such low DM mass, the dark photon portal becomes important and can dominate over the Higgs portal interaction if the mixings are not too small. However, the current constraint still can not put any bounds on $\mtn$ even for the $f/v$=3 case. Significant progress in the future could be helpful to constrain the parameter space for a light $\tn$.

\subsection{Indirect Constraints Induced from DM Annihilation}

\begin{figure}
\captionsetup{singlelinecheck = false, format= hang, justification=raggedright, font=footnotesize, labelsep=space}
\includegraphics[scale=0.6]{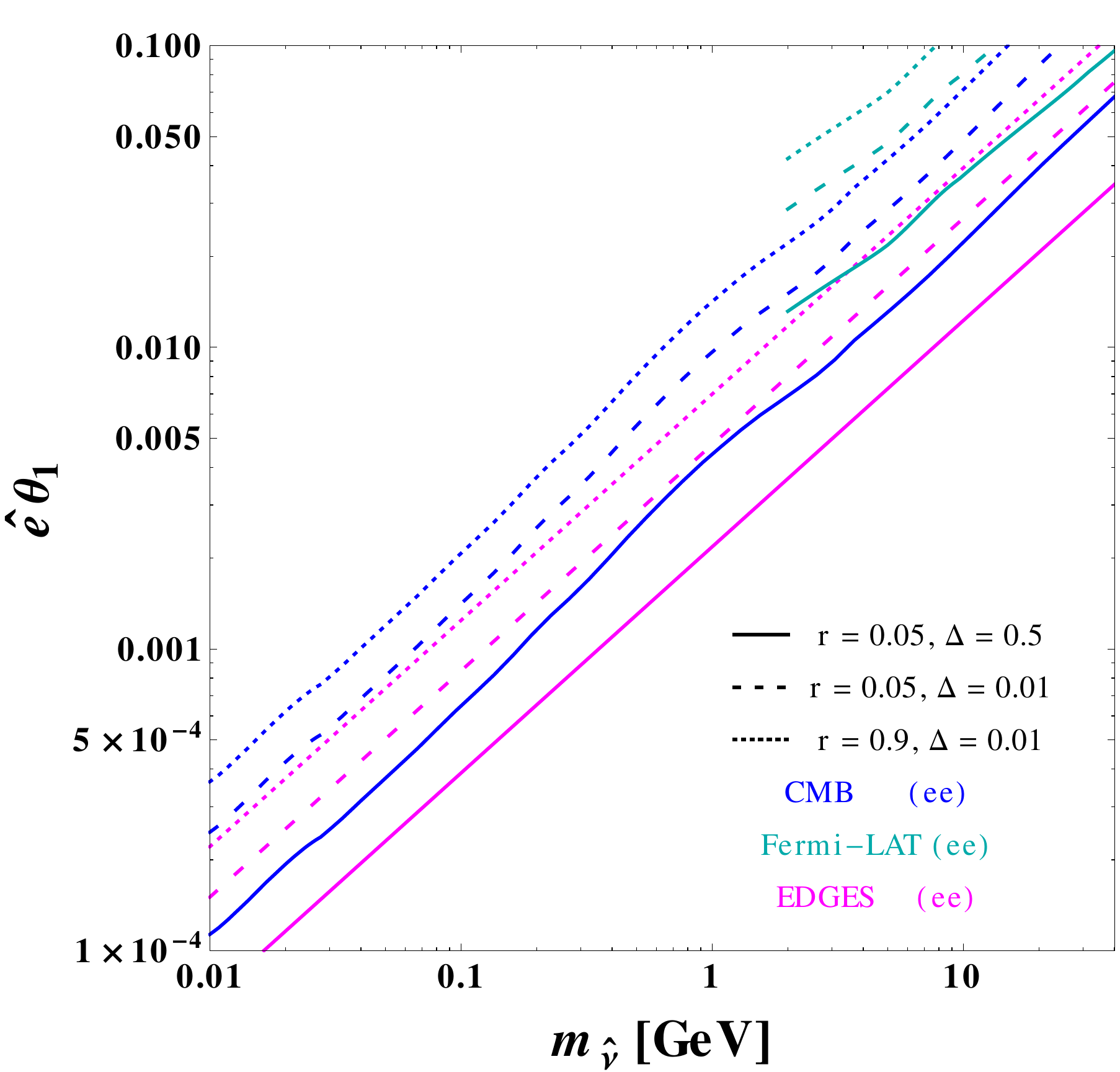}
\caption{The Fermi-LAT, CMB and EDGES bounds on DM annihilation rate in terms of $\y\theta_1$ as a function of $\mtn$. The different color curves correspond to different experiments, with the one inferred from the EDGES to be the strongest assuming that no other effect can enhance the 21cm line absorption. The solid curves represent the benchmark with $r=0.05$ and $\Delta=0.5$, where the constraints are strongest. The dashed and dotted curves are for different choices of $r$ and $\Delta$. All bounds are deduced from $ee$ final states.
}
\label{fig:ind1}
\end{figure}

Light DM is in general strongly constrained by indirect searches due to its high number density. WIMP models with annihilation cross section $\left\langle\sigma v\right\rangle \simeq 10^{-26}\text{cm}^3/\text{s}$ and $m_{\text{DM}} \lesssim10$~GeV have already been ruled out~\cite{Ade:2015xua}. In our model the DM relic density is not determined by the DM annihilation process, but by the coannihilation and coscattering processes. The DM annihilation is dominated by $\tn\tn\to\gd\gd\to 4f$, which is suppressed by $\y^4\theta_1^4$. An upper bound on the DM annihilation cross section gives a constraint on the combination of the parameters $\y \theta_1$.  Fermi-LAT data~\cite{Geringer-Sameth:2014qqa} has put an upper limit on the DM annihilation cross section for DM heavier than 6~GeV. 
A stronger constraint comes from CMB observables, which restrict the net energy deposited from DM annihilation into visible particles during the reionization era~\cite{Liu:2016cnk}. The constraints from the Fermi-LAT and the Planck data are plotted in Fig.~\ref{fig:ind1}.
 Note that the annihilation of $\tn\tn$ can produce $4e$ instead of $2e$. This may modify the bounds derived from the $2e$ final state. The total energy injection is the same, and the $4e$ final state will result in more electrons but with lower energies. Ref.~\cite{Elor:2015bho} performed a detailed study in comparing the constrains for cascade decays with different numbers of final sate particles. After convoluting with the energy dependence of the efficiency factor $f_{\text{eff}}$~\cite{Slatyer:2015jla}, it is found that the effects due to multi-step decay is rather mild.  The constraints on the $4e$ and $2e$ final states from the Planck data are roughly the same. On the other hand, a higher-step cascade tends to soften the spectrum and thus slightly weakens the constraint from the Fermi-LAT result~\cite{Elor:2015bho}. The proposed ground-based CMB Stage-4 experiment~\cite{Abazajian:2016yjj} is expected to improve the constraint by a factor of 2 to 3 compared to  Planck.
 The DM annihilation after the CMB era will also heat the intergalactic hydrogen gas and erase the absorption features of 21cm spectrum around $z\simeq 17$. The recent measurement by the EDGES experiment instead observed an even stronger absorption than the standard astrophysical expectation~\cite{Bowman:2018yin}. If one interprets this result as a constraint that the DM annihilation should not significantly reduce the absorption, the observed brightness temperature then suggests an even stronger bound than the one from CMB~\cite{DAmico:2018sxd,Cheung:2018vww,Berlin:2018sjs,Barkana:2018qrx}. In Fig.~\ref{fig:ind1} we also plot the most conservative constraint in terms of $\y\times \theta_1$ according to Ref.~\cite{DAmico:2018sxd}, taking the efficiency factor to be 1. For $\mtn =1$~GeV, the upper limit for $\y \theta_1$ is between $10^{-2}$ and $10^{-3}$ depending on other model parameters.\footnote{
The $\tn\tn$ annihilation cross section depends on the model parameters as
$$
\left\langle\sigma v\right\rangle \propto -\frac{\theta_1 ^4\y^4 \left(1-r^2\right)^{3/2}  \left(\left(\Delta ^4+4 \Delta
   ^3-8 \Delta -4\right) r^2-2 \Delta ^2 (\Delta +2)^2\right)}{32 \pi 
   (\Delta +1)^4 r^2 m_{\tn}^2 \left(\Delta ^2+2 \Delta
   -r^2+2\right)^2}.
$$
For a fixed mixing angle, the annihilation rate reaches its maximum around $\Delta\sim 0.5$ for very small $r$, while for larger $r$, it decreases as $\Delta$ increases.}
 The constraint gets more stringent for lighter $\mtn$.

\subsection{Constraints induced by the Light Twin Photon}
\label{sec:lightphoton}

In our scenario $\gd$ is the lightest twin sector particle. An on-shell $\gd$ decays through its kinetic mixing with the SM photon and there is no invisible decay mode to the twin sector. In this case, it is well described by two parameters: the twin photon mass $\mgd= r\mtn$ and the kinematic mixing with the SM photon $\epsilon$. Experimental constraints on dark photon have been extensively studied.  Summeries of current status can be found in Refs.~\cite{Curtin:2013fra,Curtin:2014cca,Alexander:2016aln,Ilten:2018crw}.

A lower bound on $\mgd$ comes from the effective number of neutrinos ($N_{\rm eff}$)~\cite{Boehm:2013jpa}. A light $\gd$ can stay in thermal equilibrium with photons and electrons after the neutrinos decouple at $T\sim 2.3$~MeV. The entropy transferred from $\gd$ to the photon bath will change the neutrino-photon temperature ratio, and therefore modify $N_{\rm eff}$. Using the results in Ref.~\cite{Boehm:2013jpa} and the Planck data~\cite{Ade:2015xua}, we obtain a lower bound on $\mgd$ around $11$~MeV. It may be further improved to $\sim 19$~MeV by the future CMB-S4 experiment~\cite{Abazajian:2016yjj}. 

\begin{figure}[h]
\captionsetup{singlelinecheck = false, format= hang, justification=raggedright, font=footnotesize, labelsep=space}
\includegraphics[scale=0.44]{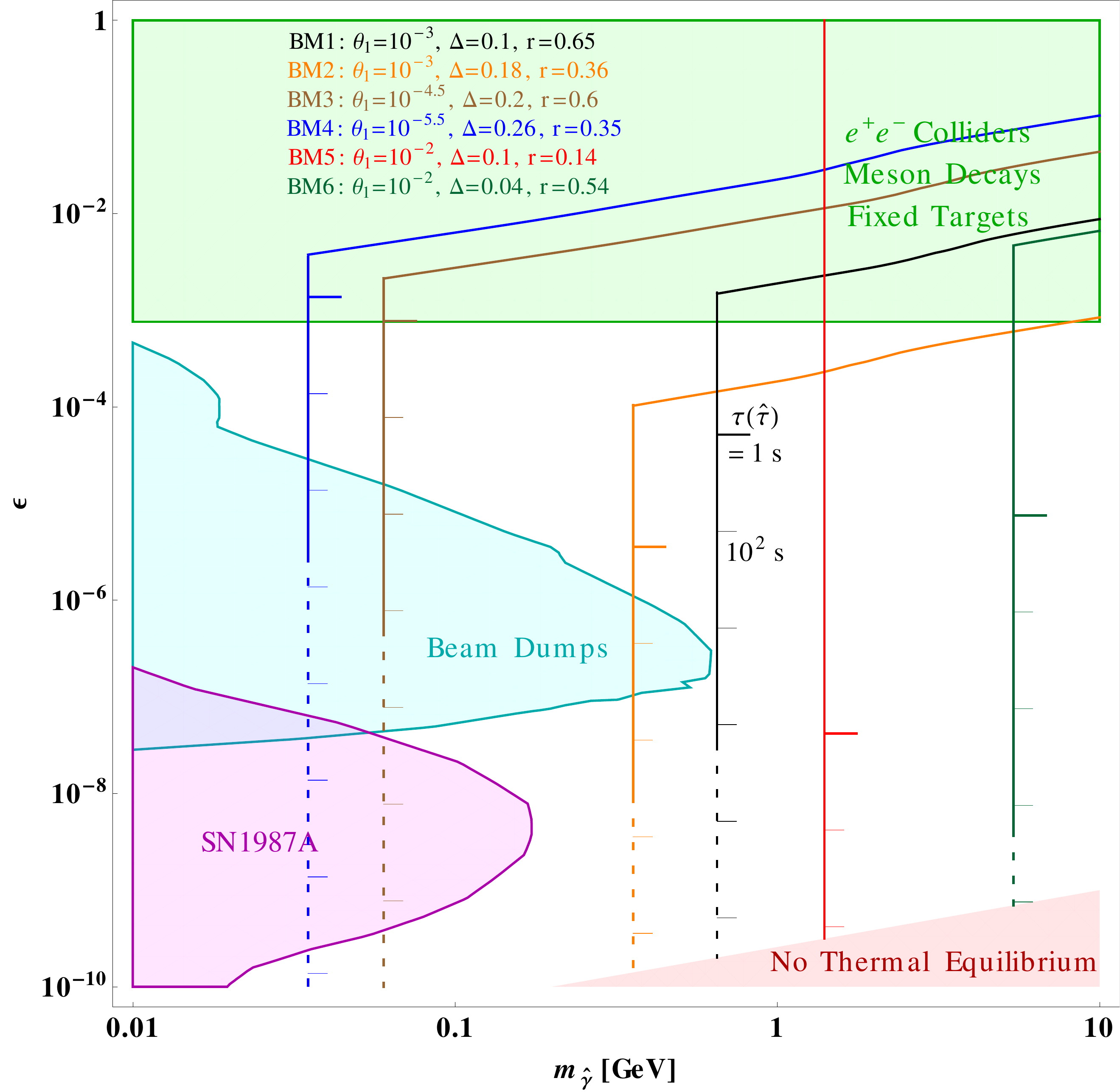}
\caption{{ Constraints on the kinetic mixing parameter $\epsilon$ and the twin photon mass $\mgd$. The green and cyan shaded regions are ruled out by lab experiments. The magenta shaded region is the constraint from SN1897A cooling. In the red shaded region $\epsilon$ is too small to keep $\gd$ in thermal equilibrium with the SM. We also plot 6 benchmark models which give the correct DM relic density from the numerical calculations in Sec.~\ref{sec:numerical}. For small enough $\epsilon$, models 1, 3, 4, and 6 are in the coscattering phase and model 2 is in the mixed phase.  At large values of $\epsilon$ these model curves turn right because the three-body (inverse) decay rate becomes large and freezes out after the coannihilation process, driving the models into the coannihilation phase. Model 5 is in the coannihilation phase for all $\epsilon$ large enough to keep $\gd$ in thermal equilibrium. The ticks on each benchmark curve represent $\tt$ lifetime, starting from $\tau(\tt)$=1~sec and increasing by $10^2$ each tick below. The dashed parts of the curves are ruled out by the BBN constraint.}}
\label{fig:darkphoton}
\end{figure}

There are also many constraints on $\epsilon$ depending on $\mgd$. 
The current upper bounds for $\epsilon$ mostly come from colliders, fixed target experiments 
and meson decay experiments, in searching for prompt decay products. (See Ref.~\cite{Ilten:2018crw} for a summary and an extended reference list.) These experiments constrain $\epsilon \lesssim 10^{-3}$ in the mass range that we consider (except for a few narrow gaps at the meson resonances). The lower bounds for $\epsilon$ come from $\gd$ displaced decays from various beam dump experiments~\cite{Bjorken:2009mm,Bergsma:1985is,
Konaka:1986cb,Riordan:1987aw,Bjorken:1988as,Bross:1989mp,
Davier:1989wz,Athanassopoulos:1997er,Astier:2001ck,
Essig:2010gu,Williams:2011qb,Blumlein:2011mv,Gninenko:2012eq,
Blumlein:2013cua}, and also from supernova SN1987A~\cite{Chang:2016ntp}. {The Big Bang Nucleosynthesis (BBN) would also constrain the lifetime of $\gd$. However, the decay of $\gd$ is only suppressed by $\epsilon$ and the BBN does not introduce extra constraints for $\epsilon \gtrsim 10^{-10}$~\cite{Chang:2016ntp}, which is required in this model to keep the DM sector in thermal contact with the SM.}
These bounds are summarized in Fig.~\ref{fig:darkphoton}.

\subsection{Constraints Induced from $\tt$ decay}
As we require $\mtn>\mgd$, the twin photon only decays to SM particles, thus $\tt$ can only be pair produced in a lab via an off-shell twin photon or from the Higgs boson decay. The constraints from $\tt$ pair production from off-shell $\gd$ are weaker compared to the ones from $\gd$ visible decay modes described in the previous subsection. Moreover, $\tn$ pair production via $\gd^*$ will be further suppressed by $\theta_1^4$, leaving $h\to \tn\tn/\tt\tt$ to be the main production channel.

At the LHC, the $\tt$ produced from $h$/$\gd^*$ will be long-lived in general, if the two-body decay $\tt \to \tn\gd$ is forbidden ($r>\Delta$), because the leading three-body decay is suppressed by $\theta_1^2\epsilon^2 \y^2 e^2$. 
Assuming $\tt$ and $\tn$ are Dirac fermions 
and taking $\mtt \simeq \mtn$, the current upper bound of the Higgs invisible decay branching ratio, Br($h \to$ invisible) $< 24\%$~\cite{Khachatryan:2016vau,Khachatryan:2016whc}, constrains $\mtn$ to be $\lesssim 19\,(52)$~GeV for $f/v=3\,(5)$.  HL-LHC is expected to improve the Higgs invisible branching ratio measurement to 6-8\%~\cite{Dawson:2013bba}, which would translate to a bound $\lesssim 11\,(30)$~GeV for $\mtn$. Future $e^+e^-$ colliders can probe BR($h\to$ invisible) to the sub-percent level~\cite{Gomez-Ceballos:2013zzn,Fujii:2015jha,CEPC-SPPCStudyGroup:2015csa,Abramowicz:2016zbo}. A 0.3\% measurement can constrain $\mtn$ down to $\lesssim 2\,(6)$~GeV for $f/v=3\,(5)$.

The $\tt$ decay width can be expressed analytically in the small $\Delta$ limit ($1\gtrsim r \gg \Delta$):
\begin{equation}
\Gamma_{\tt \to \tn e^+e^-} \simeq \frac{\theta_1^2 \Delta ^5 e^2 \y^2  \epsilon ^2}{60 \pi ^3 r^4} \mtn,
\end{equation}
which strongly depends on $\Delta$ and $r$. Numerically the proper decay length is given by
\begin{equation}
c\tau (\tt) \approx  \frac{8.8\times10^6}{N_f} \left(\frac{0.2}{\Delta}\right)^5  \left(\frac{10^{-3}}{\theta_1}\right)^2 \left(\frac{0.3}{\y}\right)^2 \left(\frac{10^{-3}}{\epsilon}\right)^2\left(\frac{r}{0.5}\right)^4 \left(\frac{1~\rm{GeV}}{\mtn}\right)\rm{cm},
\end{equation}
where $N_f$ is the number of SM fermions that can appear in the final state. 
The dark sector could be probed by searching for $\tt$ displaced decays at the HL-LHC for $c\tau(\tt)\sim \mathcal{O}$(1)~m~\cite{Curtin:2017izq}. Longer decay lengths may be tested at future proposed experiments, such as SHiP~\cite{Alekhin:2015byh},  MATHUSLA~\cite{Curtin:2017izq}, CODEX-b~\cite{Gligorov:2017nwh}, and FASER~\cite{Feng:2017uoz}.

If $\tau(\tt)\gtrsim \mathcal{O}$(1)~second, the decays of $\tt$ thermal relic will inject energy during the BBN era and the recombination era. However, the constraint is weakened by the fact that $\tt$ only makes up a small fraction of $\Omega$ after it freezes out ($\Omega_{\tt} h^2 \lesssim 10^{-3}$), also the fact that only a small fraction $\frac{\mtt-\mtn}{\mtt}=\frac{\Delta}{1+\Delta}$ of energy would be injected. The strongest bound comes from the electromagnetic decay products and depends on the model parameters. Typically, the lifetime of the long-lived $\tt$ is only constrained by BBN to be shorter than $\sim 10^6$~seconds~\cite{Poulin:2016anj,Kawasaki:2017bqm}.
The extra energy injection from $\tt$ decays could distort the CMB blackbody spectrum which can  potentially be captured by the proposed PIXIE mission~\cite{Kogut:2011xw}. It may give a slightly stronger bound than BBN on $\tau(\tt)$~\cite{Poulin:2016anj}, also around $10^{6}$~sec for our typical benchmark models.

In Fig~\ref{fig:darkphoton}, we also plot several benchmark points with fixed $r$, $\Delta$, and $\theta_1$, that give rise to the correct DM relic density from our numerical results in Sec.~\ref{sec:numerical}.  The vertical parts indicates that the thermal relic density is mostly independent of the gauge kinematic mixing parameter $\epsilon$, as long as it can keep the DM in thermal equilibrium with SM before freeze-out. For large values of $\epsilon$, the three-body (inverse) decay rate gets larger, and can even freeze out later than the coannihilation process $\hm{C_S}$. It drives the benchmark models to the coannihilation phase even if it was in the coscattering or mixed phase for smaller values of $\epsilon$. This occurs for all our benchmark models except for model 5 (red line) which is in the coannihilation phase for all $\epsilon$. The transition to the coannihilation phase reduces the DM number density, hence it needs a larger DM mass to compensate the effect. This explains the turn of the curves at lager $\epsilon$ values. However, except for model 2 (orange line), the turns occur in the region which has been ruled out by other experiments. For smaller $\epsilon$, the $\tt$ lifetime becomes longer, which is constrained by the BBN bound. The region that violates the BBN constraint is indicated by dashed lines.

\section{Conclusions}
\label{sec:conclusions}
The necessity of DM in the universe is one of the strongest evidences of new physics beyond the SM. Experimental searches in various fronts so far have not revealed the nature of the DM. For the most popular WIMP DM scenario, recent advancements in experiments have covered significant fractions of the allowed parameter space, even though there are still viable parameter space left. People have taken more seriously the possibility that DM resides in a more hidden sector, and hence has escaped our intensive experimental searches. Even in this case, it would be more satisfactory if it is part of a bigger story, rather than just arises in an isolated sector for no particular reason. In this paper, we consider DM coming from a particle in the twin sector of the fraternal twin Higgs model, which itself is motivated by the naturalness problem of the SM EW symmetry breaking and non-discovery of the colored top partners at the colliders. Although the relevant particles for the DM relic density in our study, i.e., the twin neutrino, twin tau, and twin photon, have little effect on the naturalness of the EW scale, they are an integral part of the full theory that solves the naturalness problem, just like the neutralinos in a supersymmetric SM. 

To obtain the correct DM relic density, the interplay of the twin neutrino, twin tau, and twin photon is important. The DM relic density is determined by the order of the freeze-out temperatures of various annihilation and scattering processes. It is in the coannihilation phase if the twin tau annihilation freezes out earlier than the twin neutrino to twin tau scattering.  In the opposite limit, it realizes the recently discovered coscattering phase. There is also an intermediate regime where the DM relic density is determined by both coannihilation and coscattering processes due to the momentum dependence of the coscattering process. The calculation of the DM relic density in this mixed phase is more complicated and has not been done in the literature and we provide a reasonably simple way to evaluate it with very good accuracies.

There are many experimental constraints but none of them can cover the whole parameter space. Direct detection with nuclei recoiling can only constrain heavier DM, above a few GeV. The experiments based on electron scattering are not yet sensitive to this model. Future upgrades may be able to probe the region of the parameter space with large mixings between the twin tau and the twin neutrino. Indirect constraints from DM annihilation are more sensitive to smaller DM mass with large enough twin tau -- twin neutrino mixings. Other constraints rely on the coannihilation/coscattering partners of the DM. Twin photon is subject to various dark photon constraints. Twin tau typically has a long lifetime so it is constrained by BBN and CMB. Its displaced decays may be searched at colliders with dedicated detectors or strategies. A big chunk of the parameter space still survives all the constraints. A complete coverage of the parameter space directly is not easy in the foreseeable future. An indirect test may come from the test of the whole fraternal twin Higgs model at a future high energy collider, if other heavier particles in the model can be produced.

\section*{acknowledgments}
We would like to thank Zackaria Chacko, Rafael Lang, Natalia Toro, Yuhsin Tsai, and Po-Jen Wang for useful discussions. This work is supported in part by the US Department of Energy grant DE-SC-000999. H.-C.~C. was also supported by The Ambrose Monell Foundation at the Institute for Advanced Study, Princeton.

\appendix

\section{Calculation of the Collision Operator}
\label{sec:collision}
The evaluation of the reduced collision operator $\widetilde{C}$ can be time consuming to achieve a high precision. It can be simplified by performing a partial analytic integration of the phase space, leaving a one-dimensional integral for the numerical calculation. A similar treatment for a case with a massless initial state was done in Ref.~\cite{Garny:2017rxs}. Notice that since $d\Omega_{\hm k}$ and the final state integrated $|\overline{\mathcal{M}}|^2$ in Eq.~(\ref{eqn:coredefinition}) are Lorentz invariant, $\widetilde{C}(p,t)$ can be evaluated in the rest frame of $\tn$. The initial state $\gd$ momentum $\vec{k}_r$ in the $\tn$ rest frame is a function of both $\vec{p}$ and $\vec{k}$, obtained from a simple boost. In this frame, the density distribution of $\gd$ is no longer spherically symmetric, but is still axially symmetric along the boost axis. In addition, the Mandelstam variable $s$ is a function of $|\vec{k}_r|$ only, so the integration over the angular variables can be easily performed and the result is 

\begin{equation}
\widetilde{C}=\frac{1}{2E_p}\int_{k_{rt}}^{\infty} \frac{(k_r)^2 d |\vec{k}_r|}{(2\pi)^2 2 E_{\hm k_r}}\frac{ \mtn T}{|\vec{k}_r| |\vec{p}|}\left(e^{\frac{2 |\vec{k}_r| |\vec{p}|}{\mtn T}} -1\right) e^{-\frac{E_p E_{k_r}+|\vec{k}_r| |\vec{p}| }{\mtn T}}j(s) \sigma(s),
\end{equation}
where $k_{rt}$=$\frac{1}{2 \mtn}\sqrt{(\mtt^2-\mtn^2)[(\mtt+2\mgd)^2-\mtn^2]}$ is the threshold momentum of $\gd$ for the upward scattering. 

\section{The Effect of the (Inverse) Decay}
\label{sec:ID}
When the decay $\hm{D}$ and the inverse decay $\hm{ID}$ [$\tn(p)+\gd(k) \leftrightarrow \tt(p')$] are kinematically allowed, the effect can also be included in the collision operator by an extra term $C_{\hm{ID}}$. The cross section of the inverse decay takes the form:
\begin{equation}
\sigma_{\hm{ID}}  = \frac{1}{j(s)}\int \frac{d^3 p'}{(2\pi)^3 2E_{\tt}}(2\pi)^4 \delta^4(p+k-p') |\mathcal{M}_{\hm{ID}}(s)|^2
=\frac{1}{j(s)}\frac{\pi}{\mtt}\delta(\mtt-\sqrt{s}) |\mathcal{M}_{\hm{ID}}|^2,
\end{equation}
where the flux $j(s) = 4 E_{\tn}E_{\gd}v$. The unintegrated Boltzmann equation~(\ref{eqn:c3}) now reads
\begin{equation}
Ha\partial_a f_{\tn}(q,a)=\bigg[\frac{Y_{\tt}(a)}{Y_{\tt}^{\text{eq}}(a)}f_{\tn}^{eq}(q,a)-f_{\tn}(q,a)\bigg]\bigg(\widetilde{C}_{\hm{S}}+ \widetilde{C}_{\hm{C_A}} + \widetilde{C}_{\hm{ID}}\bigg),
\end{equation}
with the inverse decay contribution given by
\begin{align}
\widetilde{C}_{\hm{ID}}(p,t)&=\frac{1}{2E_{\tn}}\int d\Omega_{\hm k} f_{\gd}^{eq}(k,t)\sigma_{\hm{ID}}(s)j(s) \\
&= \frac{1}{4\pi^2}\frac{1}{4 E_{\tn}}\frac{\pi}{\mtt}|\mathcal{M}_{\hm{ID}}|^2\int d k \frac{k^2}{E_{\gd}} f_{\gd}^{\text{eq}}(k,t)\int d \cos\theta\, \delta(\mtt-\sqrt{s})\\
&= \frac{1}{4\pi^2}\frac{1}{4 E_{\tn}}\frac{ \pi}{|\vec{p}|}|\mathcal{M}_{\hm{ID}}|^2\int_{k_{\text{min}}}^{k_{\text{max}}} d k \frac{k}{E_{\gd}} f_{\gd}^{\text{eq}}(k,t) \quad (\text{for } p\neq 0),
\end{align}
where
\begin{equation}
k_{\substack{\text{max}\\\text{min}}} = \Bigg|\frac{1}{2} \left(p \left(\Delta  (\Delta +2)-r^2\right)\pm \sqrt{\left(\mtn^2+p^2\right) \left(\Delta ^2 (\Delta +2)^2+r^4-2
		\left(\Delta ^2+2 \Delta +2\right) r^2\right)}\right)\Bigg|.
\end{equation}
Notice that the kinematic threshold of the inverse decay is lower than that of the coscattering because it does not need to create a twin photon in the final state.  For example, the ratio of the $\gd$ threshold momenta for $p=0$ is
\begin{equation}
\frac{k_T(\hm {S})}{k_T(\hm {ID})}=\frac{\sqrt{\Delta  (\Delta +2) (\Delta +2 r) (\Delta +2 r+2)}}{\left|r^2-\Delta  (\Delta +2)\right|},
\end{equation} 
which is larger than one for any $\Delta$ and $r$. The interaction rate of $\hm {S}$ is hence exponentially suppressed compared to the rate of $\hm{ID}$, besides that it has an extra $\y^2$ suppression.  
Consequently $\hm{ID}$ decouples much later than $\hm{S}$. 

To compare the decoupling time between the inverse decay and the coannihilation $\hm{C_S}$, we can simply compare $\widetilde{C}_{\hm{ID}}(p=0,t)$ with $H(t)$ when $\hm {C_S}$ decouples.
For $p=0$, $\widetilde{C}_{\hm{ID}}$ is simplified to:
\begin{align*}
\widetilde{C}_{\hm{ID}}(0,t)
= \frac{1}{4\pi^2}\frac{\pi}{2\mtn^2}|\mathcal{M}_{\hm{ID}}|^2 k f_{\gd}^{\text{eq}}(k,t).
\end{align*}
We find that in the parameter space that we consider with $\theta_1 \geq 10^{-6}$, the (inverse) decay is still active when $\hm {C_S}$ decouples. As an example, for $\theta_1=10^{-6},~\y=0.3$,~$r=0.24,~\Delta=0.26,~\mtn=0.1$~GeV, $\hm {ID}(p=0)$ decouples at $x\simeq 48$, much later than $\hm{C_S}$ that decouples at $x\simeq 30$. The difference is even larger for lower values of $\Delta$ and $r$. From these comparisons, we conclude that when $r < \Delta$, the (inverse) decay will keep $\tn$ and $\tt$ in chemical equilibrium and makes the relic density follow the coannihilation result.

\section{The Relic Density Calculation from Iteration}
\label{sec:iteration}
As the coscattering process $\hm S$ freezes out, the assumption of chemical equilibrium between $\tt$ and $\tn$ (Eq.~(\ref{eqn:CE})) breaks down. As a result, the combined Boltzmann equation for $\tt$ and $\tn$ Eq.~(\ref{eqn:COA}) no longer holds after $\hm S$ freezes out. In the mixed phase when $T_{\hm{C_S}}$ and $T_{\hm{S}}$ are comparable, in principle one should solve the coupled Boltzmann equations for both $Y_{\tn}$ and $Y_{\tt}$. As the distribution of $\tt$ remains canonical due to $\tt \gd$ scattering, we can integrate out the momentum dependence to obtain the integrated Boltzmann equation for $n_{\tt}$,
\begin{equation}
\dot{n}_{\tt}+3Hn_{\tt}=-\left\langle\sigma v\right\rangle_{C_S}(n_{\tt}^2-(n_{\tt}^{\text{eq}})^2) 
-\left\langle\sigma v\right\rangle_{C_A}(n_{\tt}n_{\tn}-n_{\tt}^{\text{eq}}n_{\tn}^{\text{eq}})-\left\langle\sigma v\right\rangle_{IS} n_{\gd}^{\text{eq}}\left(n_{\tt}-n_{\tt}^{\text{eq}}\frac{n_{\tn}}{n_{\tn}^{\text{eq}}}\right),
\label{eqn:tauBoltz}
\end{equation} 
where the last term is the contribution from the inverse coscattering process $\tt \gd \to \tn \gd$, which does not have a threshold and has very weak $p_{\tt}$ dependence. 
The combination of Eq.~(\ref{eqn:tauBoltz}) and Eq.~(\ref{eqn:intersolution}) will give the full solution of $n_{\tn}$ and $n_{\tt}$.

One way to solve the coupled equations is to use the iterative method~\cite{Garny:2017rxs}. With an initial guess of ${n_{\tn}}/{n_{\tn}^{\text{eq}}}$, we can obtain 
${n_{\tt}}/{n_{\tt}^{\text{eq}}}$ from solving Eq.~(\ref{eqn:tauBoltz}). Plugging it into Eq.~(\ref{eqn:intersolution}) will return an improved value for ${n_{\tn}}/{n_{\tn}^{\text{eq}}}$. Repeating this process will eventually converges to the exact result. 
\begin{figure}
\captionsetup{singlelinecheck = false, format= hang, justification=raggedright, font=footnotesize, labelsep=space}
\includegraphics[scale=0.3]{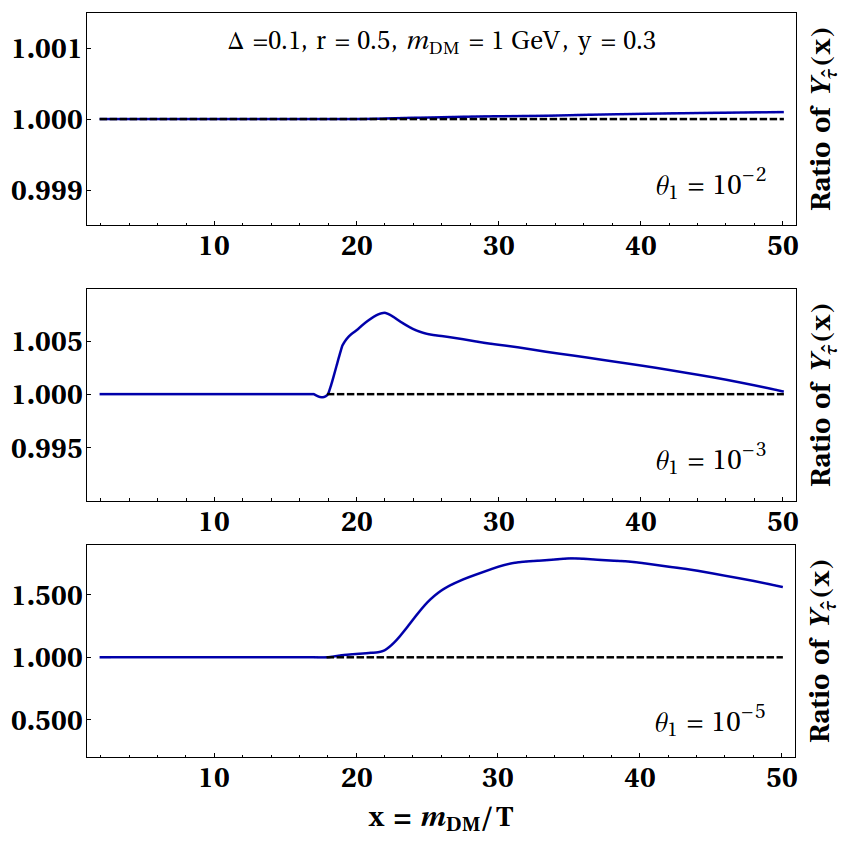}
\caption{The ratios of $Y_{\tt}$ from the first iteration and the initial input from the coannihilation calculation for 3 benchmarks differ by the mixing angle $\theta_1$ described in the text. The top plot corresponds to a coannihilation phase, where $\tn$ and $\tt$ are in approximate chemical equilibrium during the freeze-out. The bottom plot corresponds to a coscattering phase, where $\tt$ only deviates from the initial guess after $\hm S$ has already decoupled, thus  it is unable to affect $\hm S$ freeze-out. The middle plot represents a mixed phase. In this case, the maximal deviation happens simultaneously as $\hm S$ decouples, resulting in a larger iterative correction to the DM density. However, the correction is still very small.}
\label{fig:Yiteration}
\end{figure}

We take the number densities obtained from the coannihilation calculation as our starting point. We argued that in this case Eq.~(\ref{eqn:intersolution}) gives the correct results both in the coannihilation limit and the coscattering limit without iterations. The only possibility of significant deviation is when their contributions are comparable, i.e., in the mixed phase. To examine the corrections from iterations, we perform numerical studies for the benchmark parameters, $\mtn=1$~GeV, $\Delta=0.1$, $r=0.5$ and $y=0.3$, with $\theta_1= 10^{-2}, 10^{-3}, 10^{-5}$, which correspond to coannihilation, mixed, and coscattering phases respectively. We found that the correction of $n_{\tn}-n_{\tn}^{\text{eq}}$ from a single iteration is always small. The results are shown in Fig.~\ref{fig:Yiteration}. The first plot represents a coannihilation-like phase. The correction due to the coscattering contribution from the first iteration is less than $0.1\%$.   For smaller $\theta_1$ the coscattering becomes more important one can see larger deviations in $Y_{\tt}$, due to the larger $Y_{\tn}$ from decoupling of $\hm{S}$. In the coscattering phase as shown in the bottom plot, the correction of $Y_{\tt}$ at late time can be significant. However, during $\hm{S}$ freeze-out (which occurs at $x \lesssim 10$ for $\theta_1=10^{-5}$), $\tt$ is still approximately in equilibrium, and the relic density is mostly given by $n_{\tn}$. The correction to the total DM density is also small ($\mathcal{O}(10^{-6})$ in this case). The largest correction from the iteration indeed happens in the mixed phase, although it is still quite small. For $\theta_1 \simeq 10^{-3}$, the correction to $\Omega h^2$ is $\simeq 0.4\%$ from the first iteration.
The reason for such small corrections is that $Y_{\tt}$ is not as sensitive to the early $\hm S$ decoupling as $Y_{\tn}$ is, therefore using $Y_{\tt}$ from the coannihilation calculation in Eq.~(\ref{eqn:intersolution}) gives a very good approximation. Based on these results, in our numerical calculations we simply adopt this prescription without iterations.

\end{document}